\documentclass[11pt]{article}
\pdfoutput=1
\usepackage{a4wide}
\usepackage{geometry}
\usepackage{setspace}
\usepackage{indentfirst}
\usepackage[centertags]{amsmath}
\usepackage{amssymb,bm,bbm,amsfonts}
\usepackage{physics}
\usepackage{cancel}
\usepackage{bigints}
\usepackage{upgreek}
\usepackage{yfonts}
\usepackage{graphicx}
\usepackage{epstopdf}
\usepackage{caption}
\usepackage{subcaption}
\usepackage{sidecap}
\usepackage{float}
\usepackage{threeparttable}
\usepackage{rotating}
\usepackage{color}
\usepackage[dvipsnames]{xcolor}
\usepackage[normalem]{ulem}
\usepackage{verbatim}
\usepackage{tikz}
\usepackage{pgfplots}
\pgfplotsset{compat=1.16}
\usepackage[colorlinks=true,citecolor=blue,urlcolor=blue,linkcolor=blue,pdfstartview=FitH,bookmarksopen]{hyperref}
\usepackage[sort&compress,numbers]{natbib}
\numberwithin{equation}{section}

\makeatletter
\renewcommand\section{\@startsection {section}{1}{\z@}%
	{-3.5ex \@plus -1ex \@minus -.2ex}%
	{2.3ex \@plus.2ex}%
	{\normalfont\large\bfseries}}
\renewcommand\subsection{\@startsection{subsection}{2}{\z@}%
	{-3.25ex\@plus -1ex \@minus -.2ex}%
	{1.5ex \@plus .2ex}%
	{\normalfont\bfseries}}
\makeatother

\title{Cumulant dynamics in finite-memory diffusion}

\author{Navid Abbasi,$^{a}$\footnote{abbasi@lzu.edu.cn}
	\ Xin An,$^{b}$\footnote{xin.an@ugent.be}
	\ Shanjin Wu$^{c}$\footnote{shanjinwu@dlut.edu.cn}
	\\
	\\
	\small{\emph{$^{a}$School of Nuclear Science and Technology, Lanzhou University,}}\\
	\small{\emph{222 South Tianshui Road, Lanzhou 730000, China}} \\
	\small{\emph{$^{b}$Department of Physics and Astronomy, Ghent University, 9000 Ghent, Belgium}} \\
	\small{\emph{$^{c}$School of Physics, Dalian University of Technology, Dalian 116024, China}}
}

\begin{document}
	
	\setlength{\baselineskip}{16pt}
	\begin{titlepage}
		\maketitle
		
		\begin{abstract}
			Fluctuations of conserved charges are among the main proposed signatures of the quantum chromodynamics (QCD) critical point, but their interpretation requires a dynamical description of how fluctuation correlators evolve during the finite lifetime of the quark--gluon plasma (QGP) fireball.  The standard baseline for this evolution is Fickian diffusion, in which the diffusive current follows the local density gradient instantaneously. This instantaneous-current limit can miss delayed-response effects when the current-relaxation time becomes comparable to the
			relaxation time of the relevant fluctuation modes. In this work we extend this baseline to Maxwell--Cattaneo diffusion, where the current relaxes on a finite time scale and therefore retains memory. We derive closed evolution equations for multi-point Wigner functions and convert the freezeout correlators into acceptance-dependent cumulants along representative trajectories in the QCD phase diagram.
			While Fickian diffusion already causes the correlators to lag behind their instantaneous equilibrium values, finite current relaxation introduces an additional memory effect beyond this diffusive lag. As a result, current memory can suppress, shift, and reshape the
			non-monotonic behavior of the cumulants relative to both instantaneous
			equilibrium and Fickian diffusion, with the most visible effects appearing in higher-order cumulants and their ratios. These results suggest that a weak critical signal in experimentally measured cumulants need not imply the absence of a QCD critical point, but may instead reflect the suppression of the signal by memory effects in charge diffusion.
		\end{abstract}
		
		\thispagestyle{empty}
		\setcounter{page}{0}
	\end{titlepage}

	\renewcommand{\baselinestretch}{1} 
	\tableofcontents
	\renewcommand{\baselinestretch}{1.2} 
	
	\newpage
	
	\section{Introduction}
	\label{sec:intro}
	
	The evolution of a conserved charge is a classical problem in fluid mechanics
	\cite{Landau:2013fluid,Florkowski:2017olj}. A prototype problem was studied by Fourier in the context of heat conduction \cite{Fourier:1822Heat} and by Fick in the context of diffusion experiments \cite{Fick01071855}. In Fick's theory, and similarly in Fourier's theory of heat transport, an inhomogeneous distribution of a conserved charge relaxes by diffusion: on macroscopic time scales, the diffusive current is proportional to
	the spatial gradient of the local charge density. The conservation law then
	implies a global equilibration time scale \(\tau_\gamma\sim L^2/\gamma\), where
	\(L\) is the system size and \(\gamma\) is the diffusion coefficient. Notably,
	the resulting diffusion equation is parabolic and therefore permits
	instantaneous, hence acausal, propagation of arbitrarily short-wavelength
	perturbations.
	
	The Maxwell-Cattaneo (MC) theory, motivated by the earlier work of Cattaneo, Vernotte, and others on non-Fourier heat transport and non-Fickian diffusion, removes this pathology by promoting the diffusive current to an independent dynamical variable~\cite{cattaneo1958sur,Vernotte1958}. The current relaxes to its Fickian constitutive value over a finite time scale $\tau$, turning the evolution equation for the conserved charge into a hyperbolic equation with finite signal-propagation speed. In conventional interpretation, the relaxation time $\tau$ acts as a short-distance regulator. It modifies the theory at scales for which $\tau_\gamma\lesssim\tau$, while leaving the infrared hydrodynamic regime, $\tau_\gamma\gg\tau$, intact~\cite{Israel:1979wp,Heller:2020jif}.
	
	The interpretation of the current-relaxation time $\tau$ changes, however, when it represents a physical time scale rather than only a UV regulator. This can happen when $\tau$ is well separated from all other microscopic scales while remaining comparable to the macroscopic hydrodynamic time scale~\cite{Abbasi:2022aao,Abbasi:2025teu,Brants:2026qlw}, or when it effectively encodes all microscopic fluctuation dynamics through renormalization~\cite{Denicol:2024cpj,An:2025ils}. In the regime $\tau\sim\tau_\gamma$, the system can support propagating or relaxation‑type excitations associated with the dynamics of the diffusive current itself~\cite{Abbasi:2022aao,Abbasi:2025teu}. Such behavior has been experimentally realized in various systems, including electric circuits~\cite{Heaviside1876ExtraCurrent}, density fluctuations in cold atoms~\cite{doi:10.1126/science.aat4134}, and second sound in solids~\cite{PhysRev.131.2013}, and has been considered as an efficient approach to measuring transport coefficients.

	A particularly important setting in which such a finite relaxation time can
	become physically relevant is the hydrodynamic evolution near a critical
	point.  Fluctuations of the conserved charges are
	enhanced when the system is tuned to approach a critical point, and their equal-time correlation functions become dynamical
	objects whose relaxation, characterized by a relaxation time analogous of $\tau$ in MC theory, can be slow on the hydrodynamic time scale.  This
	is the essence of \emph{critical slowing down}: the fluctuation correlators do not
	instantaneously follow their local equilibrium values as the background
	fluid evolves.  They must therefore be treated as additional slow variables,
	as in Hydro+/++~\cite{Stephanov:2017ghc,An:2019csj}.  If these slow fluctuation modes are kept explicitly, one obtains an enlarged hydrodynamic description.  If instead they are integrated out, their delayed relaxation appears in the effective
	constitutive relation as memory;  the current at a given time is no
	longer determined only by the instantaneous local gradient of thermodynamic function (chemical potential per temperature unit, $\alpha=\mu/T$), but
	also depends on its relaxation history~\cite{Kovtun:2003vj,An:2019csj}. Thus, one of the essential lessons of critical dynamics is that critical fluctuations can introduce a finite relaxation into the current response, so that the current need not follow its instantaneous Fickian value.

	This provides a useful perspective on MC theory. In Hydro+/++ or related critical-dynamics frameworks \cite{Roth:2023wbp}, memory originates from additional slow fluctuation modes associated with the critical point. In MC theory, by contrast, memory is already present in the charge transport through the finite relaxation time of the diffusive current. Thus, MC theory can be regarded as a minimal laboratory for studying delayed charge transport. Rather than aiming to describe the full critical dynamics, it allows one to isolate how memory in the constitutive relation affects the evolution of conserved-charge fluctuations.
	
	These considerations connect directly to the experimental search for the quantum chromodynamics (QCD) critical point in heavy-ion collisions, where critical dynamics is probed indirectly through event-by-event fluctuations of hadron multiplicities, in particular net-proton-number fluctuations, measured over a range of collision energies. Since higher-order cumulants are more sensitive to the growth of the correlation length, they provide a sharper discriminator of critical behavior~\cite{Stephanov:2008qz,Asakawa:2009aj} --- particularly important for finite-size systems where the thermodynamic singularity is smeared~\cite{An:2025kaw}. The main idea behind the fluctuation program in heavy-ion collisions is to search for non-monotonic behavior of cumulants as functions of collision energy~\cite{Luo:2017faz}. Driven by experimental programs including the Beam Energy Scan (BES) program at RHIC and the NA61/SHINE experiment at SPS, significant progress has been made in understanding higher-order multiplicity fluctuations near the QCD critical point, both from theoretical
	and phenomenological perspectives~\cite{Koch:2008ia,Asakawa:2015ybt,Ding:2015ona,Jiang:2017mji,Stock:2018xaj,Bzdak:2019pkr,Bluhm:2020mpc,Wu:2021xgu,Dore:2022qyz,Fu:2022gou,Li:2023kja,XU:2024kpv}.

	To fully interpret precision measurements of conserved-charge fluctuations, a dynamical modeling of critical fluctuations is required~\cite{STAR:2025zdq}. A natural starting point for incorporating such nonequilibrium effects is to evolve fluctuation correlators on a time-dependent background and then convert them into acceptance-dependent cumulants at freezeout. In the absence of energy and momentum fluctuations, the baryon charge density is treated as the most relevant slow mode and its relaxation is modeled by stochastic diffusion~\cite{Ling:2013ksb,Nahrgang:2018afz,Wu:2019qfz,De:2020yyx,Nahrgang:2020yxm,Chao:2023kvz,Chattopadhyay:2023jfm}. Although the underlying dynamics is stochastic, one can derive closed deterministic evolution equations for the multi-point correlators of the charge density~\cite{An:2020vri}. These equations provide a useful  framework for quantifying how nonequilibrium evolution modifies freezeout cumulants relative to their instantaneous equilibrium values~\cite{Mukherjee:2015swa}.
	
	Pure diffusion provides the minimal nonequilibrium baseline, in which the
	charge density is evolved while the current is kept at its instantaneous
	Fickian value. Near a critical point, however, slow fluctuation modes can make the current response delayed rather than instantaneous.  In Ref.~\cite{Du:2021zqz}, this idea was implemented for the evolution of the average baryon density and diffusion current by treating the current as an additional dynamical variable with its own relaxation time.  This finite relaxation time provides an effective way to include memory in the charge-transport dynamics.
	
	In the present work, we ask how this current memory affects fluctuations
	themselves, rather than only the one-point evolution. Ref.~\cite{Sakai:2025xmo} addresses the effects of current memory on the two-point correlator near the QCD critical point. In this work, we formulate the stochastic dynamics in an enlarged variable
	space containing both the charge density and the diffusive current, and derive
	the induced evolution equations directly for the Wigner functions \(W_N\)
	associated with equal-time charge correlators. For the two-, three-, and
	four-point sectors, this yields closed equations for \(W_2\), \(W_3\), and
	\(W_4\), respectively, which reduce to the Fickian diffusion hierarchy in the
	limit \(\tau\to0\). This allows us to follow how finite current memory reshapes
	multi-point charge fluctuations before freezeout and, after acceptance
	integration, the resulting cumulants.

	This paper is organized as follows. In Sec.~\ref{sec:correlators-cumulants}, we establish the connection between equal-time correlator, expressed in terms of Wigner functions, and the acceptance-dependent cumulant in a pure diffusion setup. In Sec.~\ref{sec:evolution_correlators}, we derive closed evolution equations for the equal-time correlators in MC theory. In Sec.~\ref{sec:EoS-transport}, we specify the time-dependent thermodynamic and transport inputs along the background trajectory, using our adopted equation of state and 3D Ising mapping in the vicinity of the critical point. In Sec.~\ref{sec:numerics}, we numerically evolve the correlator equations and present the resulting modifications of freezeout spectra and acceptance-dependent cumulants, comparing throughout to the purely diffusion baseline.

	\section{Equal-time correlators and cumulants}
	\label{sec:correlators-cumulants}
	In this section, we first establish a simple relation between experimentally measured cumulants and equal-time correlators expressed in terms of Wigner functions, under the assumptions specified below.

	We work in 1+1D Minkovski coordinates $(t,z)$ where $t$ is the lab time and $z$ the longitudinal axis, while the transverse spatial coordinates are integrated out. In the literature, it is more often, however, to work with Milne coordinates $(\uptau,\eta)$, where $\uptau\equiv\sqrt{t^2-z^2}$ is the proper time and $\eta\equiv{\rm arctanh}(z/t)$ the spacetime rapidity, simply because in high-energy collisions the system is approximately boost-invariant and thus independent of rapidity $\eta$. In this work, we focus on mid-rapidity regime where the longitudinal fluid velocity $v$ remains small: $\eta\sim v=z/t\ll1$. In this regime the system can be considered approximately translationally invariant in $z$, i.e., independent of $z$. We will assume our background fulfills such properties, while we allow fluctuations to depend on both $t$ and $z$. That said, we study non-boost-invariant fluctuation dynamics on top of a Bjorken-like boost-invariant background.

\begin{figure}[t]
		\centering
		\includegraphics[width=.8\textwidth]{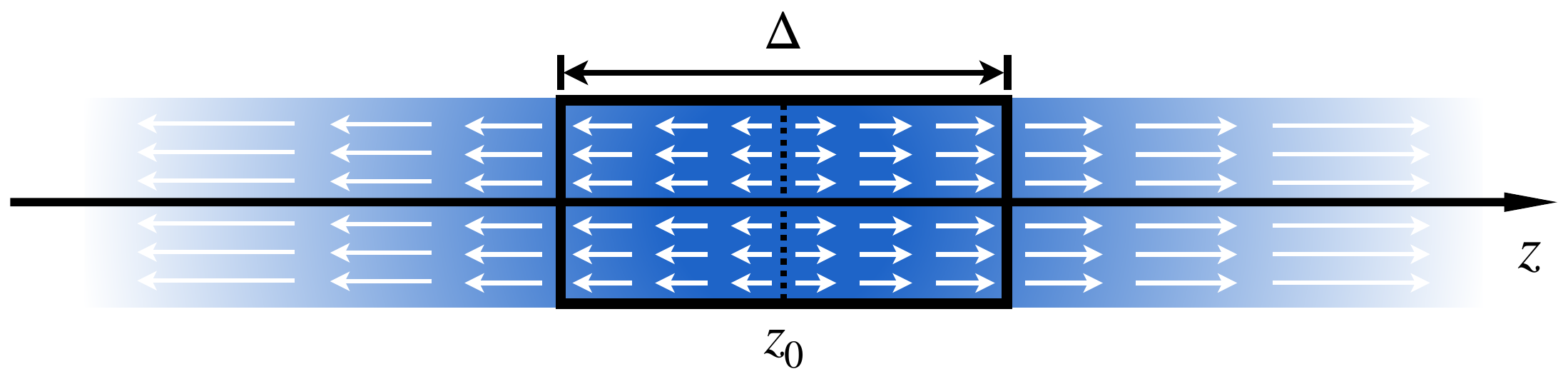}
		\caption{The mid-rapidity window in the longitudinal direction $z$ at freezeout time. The window (box in the figure) is centered around $z_0$ width $\Delta$. The vertical direction denotes the transverse coordinate that is integrated out, resulting an effective spatial 1D description. The white arrows represent the fluid/particle velocities in Bjorken flow.}\label{fig:window}
	\end{figure}

	What is directly measured in experiment, however, is the event-by-event particle multiplicity in momentum rapidity $y\equiv \operatorname{arctanh}(v_p)$, where $v_p$ is the particle velocity. For a Bjorken-like boost-invariant background, the local fluid velocity is $v=\tanh\eta$, while the emitted particles have velocities $v_p=\tanh y$ distributed around this local flow
	velocity. Therefore, near mid-rapidity, we use the approximate identification $\eta\sim y$. The measurements are limited in a finite acceptance window, which, in our problem, is characterized by a finite longitudinal interval $[z_0-\Delta/2,z_0+\Delta/2]$ centered at $z_0=0$ with width $\Delta$ (see Fig.~\ref{fig:window}). 
	
	Thus near mid-rapidity, a momentum-rapidity acceptance window $\Delta y$ corresponds approximately to a longitudinal interval $\Delta$ along the $z$-direction, i.e.,
	\begin{equation}	\label{eq:Delta}
		\Delta \;\simeq\; \uptau\,\Delta\eta\,,
		\qquad
		\Delta\eta \approx \Delta y=0.5\,,
	\end{equation}
	where ${\uptau}$ is evaluated at freezeout. This approximate identification is intended only as a minimal Bjorken-like mapping between the experimentally defined momentum-rapidity window and the longitudinal interval entering our effective description.
	Thus, in the effective spatial 1D description, the experimentally relevant cumulants are represented by integrals over a spatial interval of size $\Delta$.
	
	Within the finite acceptance window $\Delta$, the $N$-th cumulant of charge $Q$ is then defined as
	\begin{equation}\label{eq:C_N_def}
		C_N(\Delta) \equiv \langle (\delta Q)^N\rangle_c\,,  \qquad
		\delta Q(\Delta)\equiv Q(\Delta)-\langle Q(\Delta)\rangle .
	\end{equation}
	Here $Q(\Delta)$ denotes the total charge contained in the window,
	\begin{equation}
		\label{eq:Q(Delta)}
		Q(\Delta)= \int_{\Delta} dz\, n(z,t)\,, \qquad
		\int_\Delta \equiv \int_{-\Delta/2}^{\Delta/2} dz ,
	\end{equation}
	where $n(z,t)$ is the charge density per unit longitudinal length in the effective
	1D description, obtained after integrating over the transverse
	plane, so that $[n]={\rm fm}^{-1}$.
	The fluctuation of the total charge in this window is
	\begin{equation}\label{}
		\delta Q\equiv Q-\langle Q\rangle\,.
	\end{equation}
	Equivalently, 
	the  charge fluctuation in the window  can be written as
	\begin{equation}	\label{eq:dQ(Delta)}
		\delta Q(\Delta)= \int_{\Delta} dz\, \delta n(z,t)\,,
		\qquad
		\delta n\equiv n-\langle n\rangle\,.
	\end{equation}
	The \emph{connected} equal-time $N$-point function in coordinate space is defined as
	\begin{equation}\label{eq:G_N}
		G_N(z_1,\ldots,z_N;t)\equiv G_{n\dots n}(z_1,\ldots,z_N;t)\equiv
		\langle \delta n(z_1,t)\cdots \delta n(z_N,t)\rangle_c \,.
	\end{equation}
	Using Eqs.~\eqref{eq:C_N_def}, \eqref{eq:dQ(Delta)} and \eqref{eq:G_N}, the $N$-th reduced cumulant can be related to the $N$-point correlator as
	\begin{equation}\label{eq:C_N-G_N}
		C_N(\Delta)
		=
		\bigintsss_\Delta\left[\prod_{i=1}^N dz_i\right]\,
		G_N(z_1,\ldots,z_N;t) \,.
	\end{equation}

	In deterministic formulation of fluctuating hydrodynamics~\cite{An:2019csj}, we typically works with the scale separation $k\sim 1/L\ll q\sim 1/\ell$ where $L$ is the hydrodynamic inhomogeneity scale and $\ell$ is the fluctuation scale. To describe the fluctuation dynamics locally, we introduce the midpoint $\bar z$ and relative coordinates $\tilde z$:
	\begin{equation}\label{eq:midpoint}
		\bar z \equiv \frac{z_1+\cdots+z_N}{N}\,,
		\qquad
		\tilde z_i \equiv z_i-\bar z \,,
		\qquad
		\sum_{i=1}^N \tilde z_i = 0 \,.
	\end{equation}
	Following Ref.~\cite{An:2020vri}, we define the generalized Wigner function as the Fourier transform with respect to the relative coordinates $\tilde z_i$, with  conjugate fluctuation wavenumbers $q_i$:
	\begin{equation}\label{eq:W_N-G_N}
		W_N(q_1,\ldots,q_N;\bar z,t)
		\equiv{\rm WT}[G_N]\equiv
		\bigintsss \left[\prod_{i=1}^Nd\tilde z_i\, e^{-i q_i \tilde z_i}\right]
		\,\delta\!\left(\frac{1}{N}\sum_{i=1}^N \tilde z_i\right)
		\,G_N(\tilde z_1,\ldots,\tilde z_N;\bar z,t)\,,
	\end{equation}
	and vice versa,
	\begin{equation}\label{eq:G_N-W_N}
		G_N(\tilde z_1,\ldots,\tilde z_N;\bar z,t)
		=
		\bigintsss
		\left[\prod_{i=1}^N\frac{dq_i}{2\pi}\, e^{i q_i \tilde z_i}\right]
		\delta\!\left(\sum_{i=1}^N \frac{q_i}{2\pi}\right)
		W_N(q_1,\ldots,q_N;\bar z,t)\,,
	\end{equation}
	where the constraint $\sum_{i=1}^N q_i=0$ follows from global translation invariance.
	
	To derive the relation between the reduced cumulant and the Wigner function, it is useful to distinguish the full Fourier-space correlator from the local Wigner function. Using
	$\delta n(z,t)=\int \frac{dq}{2\pi}\, e^{iqz}\,\delta n_q(t)$,
	we can introduce the Fourier-space correlator
	\begin{equation}\label{eq:FW_N-G_N}
		\mathcal W_N(q_1,\ldots,q_N;t)
		\equiv
		\langle \delta n_{q_1}(t)\cdots \delta n_{q_N}(t)\rangle_c=\bigintsss  \left[\prod_{i=1}^N dz_i\, e^{-i q_i z_i}\right]\,G_N( z_1,\ldots, z_N;t)\,,
	\end{equation}
	which implies
	\begin{equation}\label{eq:G_N-FW_N}
		G_N( z_1,\ldots, z_N;t)=\bigintsss \left[\prod_{i=1}^N \frac{dq_i}{2\pi}\, e^{i q_i z_i}\right]\,\mathcal W_N(q_1,\ldots,q_N;t)\,.
	\end{equation}
	Substituting Eq.~\eqref{eq:G_N-FW_N} into Eq.~\eqref{eq:C_N-G_N} one finds
	\begin{equation}\label{eq:C_N_FW_N}
		C_N(\Delta)
		=
		\bigintsss
		\Bigg[
		\prod_{i=1}^N \frac{dq_i}{2\pi}\, A_\Delta(q_i)
		\Bigg]
		\mathcal W_N(q_1,\ldots,q_N;t)\,,
	\end{equation}
	where
	\begin{equation}\label{eq:A_Delta}
		A_\Delta(q_i)
		\equiv
		\int_\Delta dz_i\, e^{i q_i z_i}
		=
		\,\frac{2\sin(q_i\Delta/2)}{q_i}\,.
	\end{equation}
	
	The full Fourier-space correlator $\mathcal W_N$ is related to the Wigner
	function by separating the midpoint $\bar z$ from the relative coordinates
	$\tilde z_i$. Comparing their definitions in Eqs.~\eqref{eq:W_N-G_N} and \eqref{eq:FW_N-G_N} one finds
	\begin{equation}\label{eq:FW_N-W_N_int}
		\mathcal W_N(q_1,\ldots,q_N;t)
		=
		\int d\bar z\,
		e^{-i\bar z \sum_{i=1}^N q_i}\,
		W_N(q_1,\ldots,q_N;\bar z,t) \,.
	\end{equation}
	For an approximately homogeneous background, local translation invariance implies
	\begin{equation}\label{eq:trans_inv}
		W_N(q_1,\ldots,q_N;\bar z,t)\;\to\;W_N(q_1,\ldots,q_N;t)\,,
	\end{equation}
	so that the $\bar z$ integral in Eq.~\eqref{eq:FW_N-W_N_int} yields
	\begin{equation}\label{eq:FW_N-W_N}
		\mathcal W_N(q_1,\ldots,q_N;t)
		=
		\delta\!\left(\sum_{i=1}^N 
		\frac{q_i}{2\pi}\right)
		W_N(q_1,\ldots,q_N;t)\,.
	\end{equation}
	Substituting Eq.~\eqref{eq:FW_N-W_N} into Eq.~\eqref{eq:C_N_FW_N}, we finally obtain
	\begin{equation}\label{eq:C_N-W_N}
		C_N(\Delta)
		=
		\bigintsss
		\Bigg[
		\prod_{i=1}^N \frac{dq_i}{2\pi}\, A_\Delta(q_i)
		\Bigg]
		\delta\!\left(\sum_{i=1}^N \frac{q_i}{2\pi}\right)
		W_N(q_1,\ldots,q_N;t) \,.
	\end{equation}
	When $N=2$, it reproduces the expression for $C_2$ derived in Ref.~\cite{Shuryak:2000pd}.
	
	Eq.~\eqref{eq:C_N-W_N} determines the acceptance-dependent cumulants of the conserved charge and is the main result in this section. Under the the assumption in Eq.~\eqref{eq:trans_inv}, it states that the cumulants are determined by the equal-time Wigner functions evaluated on the freezeout surface and weighted by the acceptance kernel~\eqref{eq:A_Delta}. When $q\to0$, $A_\Delta(0)\to\Delta$, implying that the long-wavelength modes are almost fully accepted, while the short-wavelength modes with larger $q$ are suppressed by the average of the oscillation in the acceptence window. These $q$-dependent correlators are therefore the natural hydrodynamic inputs for acceptance-dependent observables.
	
	One should keep in mind that the conserved baryon charge often serves as a proxy for the experimentally measured (non-conserved) net-protons number, up to effects such as isospin randomization~\cite{Kitazawa:2012at}. In the present work, we do not model the final particlization stage, which would require an additional freezeout prescription for converting fluid fluctuations into particle fluctuations~\cite{Pradeep:2022mkf,Pradeep:2022eil}. Instead, we focus on the total baryon charge $Q$ measured within the acceptance window $\Delta$, as a proxy for the experimentally counted charge (net protons). Thus $C_N$ may be regarded as a proxy for experimentally measured hadronic multiplicity cumulants. 
	
	Eq.~\eqref{eq:C_N-W_N} also shows that the cumulant dynamics is controlled by the dynamics of the Wigner functions. The time evolution of $W_N$ has already been analyzed explicitly in Fickian diffusion~\cite{An:2020vri,An:2022tfk}. In the next section, we derive the evolution equations for the equal-time correlators $W_N$ with $N=2,3,4$ in MC theory.

	\section{Evolution equations for equal-time correlators}
	\label{sec:evolution_correlators}
	
	In this section we derive a closed set of evolution equations for the equal-time correlators (Wigner functions) in MC theory. As explained in Sec.~\ref{sec:correlators-cumulants}, the acceptance-dependent cumulants are determined by these correlators evaluated on the freezeout surface. 
	Our goal here is therefore to obtain dynamical equations that can be solved in a representative local fluid cell with a specified time-dependent background, and compared directly with the Fickian diffusion baseline recovered in the limit $\tau\to 0$.
	
	Throughout this section, the stochastic diffusion equations are understood as the corresponding longitudinal (1D) effective dynamics for the charge density $n(z,t)$. This effective description is valid at leading order in a gradient expansion of the slow background, with transverse structure integrated out and encoded only through the time-dependent thermodynamic and transport coefficients appearing in the evolution equations. Equivalently, one may view the correlators as evaluated in a representative local fluid cell, so that translation invariance holds locally.
	
	Our strategy is as follows. We first formulate MC dynamics as a first-order Markovian stochastic system for the density and current, $(n,J)$. In this enlarged variable space the noise is Gaussian and white, with amplitude fixed by the fluctuation--dissipation relation. We then derive the closed evolution equations for the equal-time correlators. In the main text we present the results for two-point functions explicitly, since the two-point sector already displays the basic structure of the formalism and its Fickian diffusion limit; the corresponding results for higher-point functions are collected in Appendix~\ref{app:numerical}.
	
	\subsection{Maxwell-Cattaneo stochastic dynamics}
	\label{subsec:MC_dynamics}
	
	The MC stochastic equations for conserved density $n$ and diffusive current $J_i$ read\footnote{The MC diffusion problem can be extended by including the ideal part of the current. We do not consider this extension in the current work~\cite{CHRISTOV2009481}.}
	\begin{subequations}\label{eq:MC}
		\begin{align}
			\partial_t n + \nabla\!\cdot J = &\,0\,,\label{eq:MC_n}\\
			\tau\,\partial_t J_i + J_i =& -\lambda\,\nabla_i \alpha+ \sqrt{\sigma T}\,\eta_i\,.\label{eq:MC_J}	
		\end{align}
	\end{subequations}
	Here $\tau$ is a relaxation time over which the current $J_i$ approaches its Fickian limit: when $\tau\to 0$, Eq.~\eqref{eq:MC_J} reduces to the constitutive relation in Fickian diffusion.
	The coefficients $\lambda$ and $\sigma$ are thermal and electrical conductivities respectively, $\alpha=\mu/T$ is the chemical
	potential in the unit of temperature, and $\eta_i$ is a Gaussian white noise satisfying the local correlation
	\begin{equation}\label{eq:eta-eta}
		\langle \eta_i(x_1,t_1)\,\eta_j(x_2,t_2) \rangle
		= 2\,\delta_{ij}\,\delta^{(3)}(x_1-x_2)\,\delta(t_1-t_2)\, .
	\end{equation}
	Since the diffusive current vanishes in equilibrium, the thermodynamic and transport
	coefficients are taken to depend only on the local density $n$. In a diffusion problem, one also treats temperature $T$ (or energy) as a time-dependent background field. 
	Thus, we have $\tau=\tau(n)$, $\lambda=\lambda(n)$, $\sigma=\sigma(n)$.
	
	The fluctuation-dissipation relation (or KMS condition~\cite{Liu:2018kfw}) implies the following constraints on the above thermodynamic functions and transport coefficients:
	\begin{equation}	\label{eq:KMS}
		\lambda=T \sigma\,, \quad (\log\tau)'=(\log\lambda)'\,,
	\end{equation}
	where in this work we specifically use $(\dots)'\equiv\delta(\dots)/\delta n$.  A detailed derivation of Eq.~\eqref{eq:KMS} is presented in Appendix~\ref{app:KMS}. The first identity in Eq.~\eqref{eq:KMS} is known as the fluctuation--dissipation relation  between the noise amplitude to the transport coefficient, while the second identity imposes nontrivial constraint on $\tau(n)$. In this work we focus on the simplified case in which $\lambda$ and $\tau$ are both constants, being consistent with Eq.~\eqref{eq:KMS}. As a consequence, the nonlinearity enters solely through the equation of state $\alpha=\alpha(n)$ where $n$ is the fluctuating density field.  Expanding around averaged field $\langle n \rangle$, we obtain
	\begin{equation}
		\label{eq:alpha_expansion}
		\alpha(n)=\alpha(\langle n\rangle)+\alpha' \delta n+\frac12 \alpha'' \delta n^2+\frac{1}{6}\alpha'''\delta n^3+\cdots.
	\end{equation}
The nonlinearity of $\alpha(n)$ generates the couplings to correlation functions $G_n$ through averaging the stochastic equation~\eqref{eq:alpha_expansion}:
        \begin{equation}
        \langle\alpha(n)\rangle
        =\alpha(\langle n\rangle)
        +\frac{1}{2}\alpha''G_2
        +\frac{1}{6}\alpha'''G_3+\cdots\,,
        \end{equation}
If the averaged equations were iterated self-consistently, these nonlinear
terms would give rise to the dynamical feedback of the correlators on the one-point function. This fluctuation feedback is a higher-order effect in
the power counting adopted in Sec.~\ref{subsec:truncation} and can therefore be consistently neglected at leading order. Near the critical regime, however, the correlators relax on long time scales, so that their response to the evolving background is delayed and their feedback induces memory in the effective one-point dynamics~\cite{An:2019csj}. In the present work, rather than evolving this feedback explicitly, we model the associated memory at the level of the transport sector by promoting the dissipative current to an independent dynamical variable with a finite relaxation time $\tau$. The Maxwell--Cattaneo equation thus provides an effective single-relaxation-time description of the corresponding nonlocal-in-time current response.

	To put the dynamics in a convenient Markovian form in Fourier space, we introduce the ``generalized momentum''
	\begin{equation}\label{eq:g_q}
		g\equiv \partial_t n\,,
	\end{equation}
	which couples only to the longitudinal component of the current. 
	We then obtain a dynamics system with two scalar variables, $(n,g)$. The resulting nonlinear one-point equations provide a starting point for deriving closed evolution
	equations for equal-time multi-point Wigner functions by taking time derivatives and using the product rule together with the equal-time It\^{o} prescription:
	\begin{subequations}\label{eq:MC_1pt}
		\begin{align}
			\partial_t\delta n &= \delta g\,, \label{MC_1pt_n}\\
			\tau \partial_t \delta g &=\, - \delta g +\gamma \nabla^2 \delta n 
			+\frac{1}{2} \gamma'\nabla^2 (\delta n)^2
			+\frac{1}{6} \gamma''\nabla^2  (\delta n)^3 +\dots+ \zeta\,,
			\label{MC_1pt_g}
		\end{align}
	\end{subequations}
	where $\gamma\equiv \lambda\alpha'$  and $\zeta\equiv -\nabla\!\cdot(\sqrt{\lambda}\,\eta)$. The expansion in fluctuation field $\delta n$ is truncated at third order, for the sake of deriving equations for four- and lower-point functions. Introducing the corresponding Fourier modes
	\begin{equation}\label{eq:Fourier_modes}
		\delta n_{\mathbf q}(t) =\int_\mathbf x\, e^{-i\mathbf q\cdot\mathbf x}\,\delta n(\mathbf x,t)\,, \quad \delta g_{\mathbf q}(t) =\int_\mathbf x\, e^{-i\mathbf q\cdot\mathbf x}\,\delta g(\mathbf x,t)\,, \quad   \zeta_{\mathbf q}(t) =\int_\mathbf x\, e^{-i\mathbf q\cdot\mathbf x}\,\zeta (\mathbf x,t)\,, 
	\end{equation}
	one can also write these equations as
	\begin{subequations}\label{eq:MC_1pt(q)}
		\begin{align}
			\partial_t\delta n_{\mathbf q} &= \delta g_{\mathbf q}\,, \label{MC_1pt_n(q)}\\
			\tau \partial_t \delta g_{\mathbf q} &=\,: - \delta g_{\mathbf q} - \gamma q^2 \delta n_{\mathbf q} 
			- \frac{1}{2} \gamma'q^2 \!\int_{\mathbf k}\delta n_{\mathbf k} \delta n_{\mathbf q-\mathbf k}
			- \frac{1}{6} \gamma''q^2 \!\int_{\mathbf k,\mathbf l} \delta n_{\mathbf k} \delta n_{\mathbf l} \delta n_{\mathbf q-\mathbf k-\mathbf l} + \zeta_{\mathbf q}:\,,
			\label{MC_1pt_g(q)}
		\end{align}
	\end{subequations}
	where
	\begin{equation}		\label{eq:theta-theta}
		\langle \zeta_{\mathbf{q}_1}(t)\,\zeta_{\mathbf{q}_2}(t')\rangle
		= -2\lambda\,\mathbf{q}_1\cdot \mathbf{q}_2 (2\pi)^3 \delta^{(3)}(\mathbf{q}_1+\mathbf{q}_2)\,\delta(t_1-t_2)\,.
	\end{equation}
	The derivation of the evolution equations for correlators follows from the nonlinear equations for one-point functions, \eqref{eq:MC_1pt} or~\eqref{eq:MC_1pt(q)}~\cite{An:2020vri}.

	\subsection{Scale separation and power counting}
	\label{subsec:truncation}
	
	Before discussing the evolution of multi-point functions, it is useful to summarize  the scale separation underlying our formalism and the power counting that justifies the truncation of the correlator hierarchy.
	
	Fluctuations are present on all length scales, but their relative importance
	is often suppressed by the number $\mathcal N$ of effective degrees of freedom, such as colors in gauge theory, particles in gas, or correlated cells in fluid. Following Ref.~\cite{An:2020vri,An:2022jgc}, we introduce a small parameter $\epsilon\sim1/\mathcal N$ that controls the magnitude of fluctuations. For connected correlators one has
	\begin{equation}	\label{eq:powercounting_G}
		G^c_{i_1\cdots i_n}\sim \epsilon^{\,N-1}\,,
	\end{equation}
	since, heuristically, $N$-point connected correlator requires $N-1$ statistical connections. The truncation is then controlled by the power counting in Eq.~\eqref{eq:powercounting_G}: the two-point sector $G_2$ closes on itself at $O(\epsilon)$, the three-point sector closes on $(G_2,G_3)$ at $O(\epsilon^2)$, and the connected four-point sector closes on $(G_2,G_3,G_4)$ at $O(\epsilon^3)$.
	
	The same power counting also clarifies the approximation adopted following Eq.~\eqref{eq:alpha_expansion}.  A fluctuation-induced correction to the one-point thermodynamic
	sector first arises through $G_2\sim O(\epsilon)$. Feeding such a correction back into the correlator hierarchy would therefore modify the leading evolution retained in each multi-point sector only at the next order in $\epsilon$.  We consequently evolve the correlators on the prescribed local equilibrium background and neglect their backreaction through the equation of state.
	
	When the system is close to, but not exactly in, local equilibrium, another small parameter appears, $\epsilon_q \sim q\,\ell_{\rm mic}$. Here $\ell_{\rm mic}$ is the microscopic scale, which is much smaller than the macroscopic fluctuation scale $1/q$. 
	The parameter $\epsilon_q$ controls the validity of the gradient expansion,
	but it does not by itself determine whether fluctuation modes can be treated
	as equilibrated.

	For this purpose one must compare their equilibration time
	with the macroscopic evolution time of the background.  We denote the latter by $\tau_h$, which in our setup is set by the Bjorken expansion time.  In the hydro-kinetic description~\cite{Akamatsu:2016llw,An:2019osr}, the equilibration time of a fluctuation mode with wavenumber $q$ is $\tau_\gamma(q)\sim 1/\gamma q^2.$ Harder modes relax faster and can therefore be treated as locally
	equilibrated on the hydrodynamic time scale, while softer modes relax more slowly, and when $\tau_\gamma(q)\gtrsim \tau_h$, their dynamics cannot be replaced by instantaneous equilibrium.  However, the feedback contribution of very soft modes is phase-space suppressed~\cite{Akamatsu:2016llw}.  The dominant feedback therefore
	comes from an intermediate hydro-kinetic window of modes with
	$\tau_\gamma(q)\sim\tau_h$. These modes are slow enough that their
	equilibration is not instantaneous, but not so soft that their phase space is
	negligible.

	This hydro-kinetic window is particularly relevant in heavy-ion collisions, where the QGP expands and hadronizes over a short time scale, $\tau_h\sim 1$--$10\,{\rm fm}$.  Although the bulk medium may
	hydrodynamize rapidly, this does not imply that the fluctuation correlators instantaneously track their local-equilibrium values throughout the subsequent evolution~\cite{Heller:2015dha}. Modes with $\tau_\gamma(q)\sim\tau_h$ can still relax on time scales
	comparable to the hydrodynamic evolution time, making their explicit
	dynamics potentially relevant. Far from the critical point, however,  this effect is not expected to produce a substantial modification of the freezeout cumulants relative to their instantaneous local-equilibrium values. In this case, the MC equation with a non-critical relaxation time provides a useful baseline for critical effects.

	Near the critical point, however, this hydro-kinetic window is enlarged by  critical slowing down. The transport coefficient $\gamma$ is suppressed by the growing correlation length $\xi$: $\gamma\sim\xi^{-a}$, with $a=2$ in Model B and $a=1$ in Model H~\cite{Hohenberg:1977ym}. Consequently, for typical critical modes with $q\sim\xi^{-1}$, one has $\tau_\gamma(q\sim\xi^{-1})\sim\xi^{2+a}\gg\tau_h$.
	Modes that would be treated as locally equilibrated away from critical point may therefore become dynamical near the critical region.  This is the basic motivation behind Hydro+/++~\cite{Stephanov:2017ghc,An:2019csj}. The explicitly evolved correlators can then deviate substantially from their instantaneous equilibrium values on the freezeout surface.

	These critically slowed fluctuation modes in turn produce a delayed feedback on the hydrodynamic variables, in particular on the current $J^\mu$. According to Ref.~\cite{An:2019csj}, the slowest mode contributing to this feedback is the charge--momentum correlator.  In our effective description, the delayed current response is modeled by the MC relaxation time $\tau$ of the diffusive current. 
	
	This memory effect becomes relevant when $\tau$ is comparable to the relaxation time of the fluctuation modes.  In the diffusive regime, a mode with wavenumber $q$ relaxes on the scale $\tau_{\gamma}\sim 1/\gamma q^2\,.$ The competition between this diffusive relaxation time and the current memory time defines the MC characteristic wavenumber
	\begin{equation}	\label{eq:powercounting_qstar}
		q_\star = \frac{1}{2\sqrt{\tau\,\gamma}}\,.
	\end{equation}
	For modes with $q\ll q_\star$, the current relaxes rapidly on the diffusive time scale and the Fickian limit is a good approximation.  As $q$ approaches $q_\star$, finite-current-relaxation effects become important, and the instantaneous Fickian description is no longer reliable.   We discuss their impact in different scenarios below.

	\subsection{Evolution equations for two-point functions}
	\label{subsec:W2}

	In this subsection we present the closed evolution equations for the two-point
sector. Following the Wigner transform defined in Eq.~\eqref{eq:W_N-G_N}, we introduce
	\begin{equation}\label{eq:def_WXY2}
		W_2(q)\equiv {\rm WT}[ G_{nn}]\,,\qquad
		X_2(q)\equiv{\rm WT}[ G_{gn}]\,,\qquad
		Y_2(q)\equiv{\rm WT}[ G_{gg}]\,,
	\end{equation}
	where $G_{nn}(z_1,z_2;t)\equiv\langle\delta n(z_1;t)\delta n(z_2;t)\rangle$ was defined inn Eq.~\eqref{eq:G_N}, and 
	\begin{equation}\label{eq:def_G2}
		G_{gn}(z_1,z_2;t)\equiv\langle\delta g(z_1;t)\delta n(z_2;t)\rangle\,,\qquad
		G_{gg}(z_1,z_2;t)\equiv\langle\delta g(z_1;t)\delta g(z_2;t)\rangle\,.
	\end{equation}
	Here $g$ is the auxiliary field introduced in Eq.~\eqref{eq:g_q}. Thus \(W_2\),
	\(X_2\), and \(Y_2\) denote equal-time Wigner functions with zero, one, and two
	insertions of the auxiliary field \(g\), respectively. The physical charge
	correlator is \(W_2\), while \(X_2\) and \(Y_2\) are the  mixed charge--auxiliary and the  auxiliary correlators
	generated by the Markovian completion of the MC dynamics.

	At leading Gaussian order, the nonlinear couplings $\gamma'$ and $\gamma''$ do not contribute to the closed two-point sector. Retaining only the terms linear in the fluctuating fields, Eqs.~\eqref{MC_1pt_n} and \eqref{MC_1pt_g} reduce to
\begin{subequations}
\begin{align}
		d\delta n_{\mathbf  q}
		&=
		\delta g_{\mathbf  q}\,dt,
		\label{eq:linear-n-differential}
		\\
		d\delta g_{\mathbf  q}
		&=
		-\frac{1}{\tau}
		\left(
		\delta g_{\mathbf  q}
		+\gamma q^2\delta n_{\mathbf  q}
		\right)dt
		+\frac{1}{\tau}d\Xi_{\mathbf  q},
		\label{eq:linear-g-differential}
	\end{align}
\end{subequations}
where $d\Xi_{\bm q}\equiv \zeta_{\bm q}\,dt$ denotes the stochastic increment. From Eq.~\eqref{eq:theta-theta}, its It\^{o} contraction is \footnote{We use the equal-time It\^o prescription, in which the stochastic increment
is uncorrelated with fields evaluated immediately before the
increment~\cite{Gardiner:2009}.}
	\begin{equation}
		\left\langle
		d\Xi_{\mathbf q_1}d\Xi_{\mathbf q_2}
		\right\rangle
		=	-2\lambda\,\mathbf q_1\cdot\mathbf q_2\,
		(2\pi)^3\delta^{(3)}(\mathbf q_1+\mathbf q_2)\,dt .
		\label{eq:noise-increment}
	\end{equation}

	We first apply the product rule to the density correlator:
	\begin{align}
		\partial_t
		\left\langle
		\delta n_{\mathbf  q}\delta n_{-\mathbf  q}
		\right\rangle
		&=
		\left\langle
		\delta g_{\mathbf  q}\delta n_{-\mathbf q}
		\right\rangle
		+
		\left\langle
		\delta n_{\mathbf q}\delta g_{-\mathbf q}
		\right\rangle .
		\label{eq:W2-intermediate}
	\end{align}
	Since the fields are real,
	$\delta n_{\mathbf q}^{*}=\delta n_{-\mathbf q}$ and
	$\delta g_{\mathbf q}^{*}=\delta g_{-\mathbf q}$. Consequently, the two terms on
	the right-hand side are complex conjugates, and their Wigner transforms give
	$X_2(q)+X_2^{*}(q)=2\operatorname{Re}X_2(q)$.
	
	For the mixed correlator, the It\^{o} product rule gives
	\begin{align}
		d\left(
		\delta g_{\mathbf q}\delta n_{-\mathbf q}
		\right)
		&=
		\delta n_{-\mathbf q}\,d\delta g_{\mathbf q}
		+
		\delta g_{\mathbf q}\,d\delta n_{-\mathbf q},
		\label{eq:X2-product}
	\end{align}
	where there is no quadratic It\^{o} term $d\delta g_{\mathbf q}d\delta n_{-\mathbf q}$ because $d\delta n_{\bm q}$
	contains no stochastic increment (see Eq.~\eqref{eq:linear-n-differential}). Substituting
	Eqs.~\eqref{eq:linear-n-differential} and
	\eqref{eq:linear-g-differential}, and using the fact that an It\^{o}
	stochastic increment has zero correlation with fields evaluated at the same
	time before the increment, one finds
	\begin{align}
		\tau\partial_t
		\left\langle
		\delta g_{\mathbf q}\delta n_{-\mathbf q}
		\right\rangle
		&=
		-
		\left\langle
		\delta g_{\mathbf q}\delta n_{-\mathbf q}
		\right\rangle
		-\gamma q^2
		\left\langle
		\delta n_{\mathbf q}\delta n_{-\mathbf q}
		\right\rangle
		+\tau
		\left\langle
		\delta g_{\mathbf q}\delta g_{-\mathbf q}
		\right\rangle .
		\label{eq:X2-intermediate}
	\end{align}
	
	On the other hand, for the auxiliary-field correlator one must retain the quadratic It\^{o} contribution \footnote{In contrast to ordinary deterministic calculus, quadratic stochastic increments cannot be discarded; since $dW\sim\sqrt{dt}$, one has $(dW)^2\sim dt$. Thus the It\^o product rule reads $d(XY)=X\,dY+Y\,dX+dX\,dY$, or $d(X^2)=2X\,dX+(dX)^2$.}:
	\begin{align}
		d\left(
		\delta g_{\mathbf q}\delta g_{-\mathbf q}
		\right)
		={}&
		\delta g_{-\mathbf q}\,d\delta g_{\mathbf q}
		+
		\delta g_{\mathbf q}\,d\delta g_{-\mathbf q}
		+
		d\delta g_{\mathbf q}\,d\delta g_{-\mathbf q}.
		\label{eq:Y2-product}
	\end{align}
	Using Eq.~\eqref{eq:noise-increment} with
	$\mathbf q_2=-\mathbf  q_1=-\mathbf  q$, the last term becomes
	\begin{equation}
		\left\langle
		d\delta g_{\mathbf q}\,d\delta g_{-\mathbf q}
		\right\rangle
		=	\frac{2\lambda q^2}{\tau^2}\,dt .
		\label{eq:Y2-Ito-source}
	\end{equation}
	We therefore obtain
	\begin{align}
		\tau\partial_t
		\left\langle
		\delta g_{\mathbf q}\delta g_{-\mathbf q}
		\right\rangle
		={}&
		-2
		\left\langle
		\delta g_{\mathbf q}\delta g_{-\mathbf q}
		\right\rangle
		-\gamma q^2
		\left[
		\left\langle
		\delta g_{\mathbf q}\delta n_{-\mathbf q}
		\right\rangle
		+
		\left\langle
		\delta n_{\mathbf q}\delta g_{-\mathbf q}
		\right\rangle
		\right]
		+\frac{2\lambda q^2}{\tau}.
		\label{eq:Y2-intermediate}
	\end{align}
	
	Taking the Wigner transform and keeping the leading term in the gradient expansion then yields
	\begin{subequations}
		\label{eq:two-point-evolution}
		\begin{align}
			\partial_t W_2(q)
			&=
			2\operatorname{Re}X_2(q),
			\label{eq:evo_W2}
			\\
			\tau\partial_t X_2(q)
			&=
			-X_2(q)-\gamma q^2W_2(q)+\tau Y_2(q),
			\label{eq:evo_X2}
			\\
			\tau\partial_t Y_2(q)
			&=
			-2Y_2(q)
			-2\gamma q^2\operatorname{Re}X_2(q)
			+\frac{2\lambda q^2}{\tau}.
			\label{eq:evo_Y2}
		\end{align}
	\end{subequations}
	The last term in Eq.~\eqref{eq:evo_Y2} is the equal-time noise source required by the fluctuation--dissipation relation. Indeed, in a stationary  equilibrium background, Eqs.~\eqref{eq:two-point-evolution} give
	\begin{equation}
	X_2^{\rm eq}=0,\qquad
	Y_2^{\rm eq}=\frac{\lambda q^2}{\tau},\qquad
	W_2^{\rm eq}=\frac{\lambda}{\gamma}
	=\frac{1}{\alpha'},
	\end{equation}
	where $\gamma=\lambda\alpha'$ has been used.

	We now show that Eqs.~\eqref{eq:evo_W2}--\eqref{eq:evo_Y2} reduce to the Fickian diffusion problem in the limit $\tau\to0$. This limit has to be taken by eliminating
	 \(X_2\) and \(Y_2\) as fast fields.  Equivalently,
	one drops the time-derivative terms in Eqs.~\eqref{eq:evo_X2} and~\eqref{eq:evo_Y2},  while keeping the finite contribution of \(\tau Y_2\).  This gives the algebraic constraints
	\begin{subequations}\label{eq:XY2_equil}
		\begin{align}
			-X_2(q)-\gamma q^2 W_2(q)+\tau\,Y_2(q)&=0\,,\\
			-2Y_2(q)-2\gamma q^2\,\mathrm{Re}\,X_2(q)+\frac{2\lambda}{\tau}\,q^2&=0\,.
		\end{align} 
	\end{subequations}
	Although \(Y_2\sim \lambda q^2/\tau\) is singular in this limit, the
combination \(\tau Y_2\) remains finite.  Solving the above constraints gives
	\begin{equation}\label{eq:X2(q)_Fick}
		X_2(q)=-\gamma q^2 W_2(q)+\lambda q^2\,.
	\end{equation}
	Substituting this result into Eq.~\eqref{eq:evo_W2}, and using 	\(W_2^{\rm eq}=\lambda/\gamma=1/\alpha'\) and
	\({\rm Im}\,W_2(q)=0\), one obtains
	\begin{equation}\label{W_2}
		\partial_t W_2(q)=-2\gamma q^2\bigl(W_2(q)-W_2^{\rm eq}\bigr)\,,
	\end{equation}
		This is precisely the relaxation equation for the equal-time two-point
	correlator in the Fickian diffusion problem~\cite{An:2022tfk}.  Thus the
	\(\tau\to0\) limit is singular only in the auxiliary sector, while the physical
	charge correlator \(W_2\) remains finite and smoothly approaches the standard
	diffusive result.

The decrease of the relaxation time with increasing $q$ has a simple
real-space interpretation. A fluctuation mode with wavenumber $q$ consists
of neighboring regions of excess and deficit charge separated by a distance
of order $\ell\sim q^{-1}$. Diffusion relaxes the mode by transporting charge between
these regions. Since the characteristic diffusion distance satisfies
$\ell^2\sim\gamma t$, redistributing charge over a distance
$\ell$ requires a time scale of order $t\sim1/\gamma q^2$. Thus shorter-wavelength modes relax more rapidly because charge is
transported over a shorter distance. Eq.~\eqref{W_2} describes the relaxation of two-point
function, for which the two density insertions contribute additively to the
decay rate $1/2\gamma q^2$. This estimate applies in the long-wavelength Fickian regime. As $q$
approaches $q_\star$ (see Eq.~\eqref{eq:powercounting_qstar}), the finite relaxation time of the current becomes
important, and the simple $q^{-2}$ decrease behavior should no longer be extrapolated.

	The same procedure applies to the higher-point sectors.  The explicit evolution
	equations for the three- and four-point Wigner functions are lengthy and are
	therefore collected in Appendix.~\ref{app:numerical}, together with details of the numerical
	implementation.

	\section{Equation of state and transport coefficients}
	\label{sec:EoS-transport}
	In Sec.~\ref{sec:evolution_correlators} we have derived a closed set of evolution equations for the equal-time Wigner functions in MC theory.  To perform phenomenological simulations,
	we must now specify the time-dependent input functions appearing in those equations, namely the thermodynamic derivatives (equation of state) encoded in $\alpha(n(t))$ and the transport coefficients $\gamma(t)$ (and, in MC theory, $\tau(t)$).  In this section we summarize our parameterization of the baryon susceptibilities and diffusion coefficient, based on a mapping to the 3D Ising
	universality class.
	
	Note that the evolution equations solved in this work are the longitudinal (1D) effective dynamics for charge fluctuations, appropriate when transverse structure is integrated out and the background varies slowly on the scales of interest.
	The mapping to the 3D Ising model is used only to parameterize the \emph{local static}
	thermodynamic input near the critical point, namely the susceptibilities $\chi_n(T,\mu)$ along the background trajectory.
	In other words, ``3D Ising'' specifies the universality class of equilibrium fluctuations, while ``1D'' refers to the reduced spacetime dependence of the dynamical evolution we simulate; these two statements refer to different aspects of the setup and are therefore compatible.

	The transverse area and any residual transverse dependence enter only through overall normalization and
	through the time-dependent input functions, and are not evolved explicitly in the present 1D setup.

\begin{figure}[t]
		\centering
		\includegraphics[width=.75\textwidth]{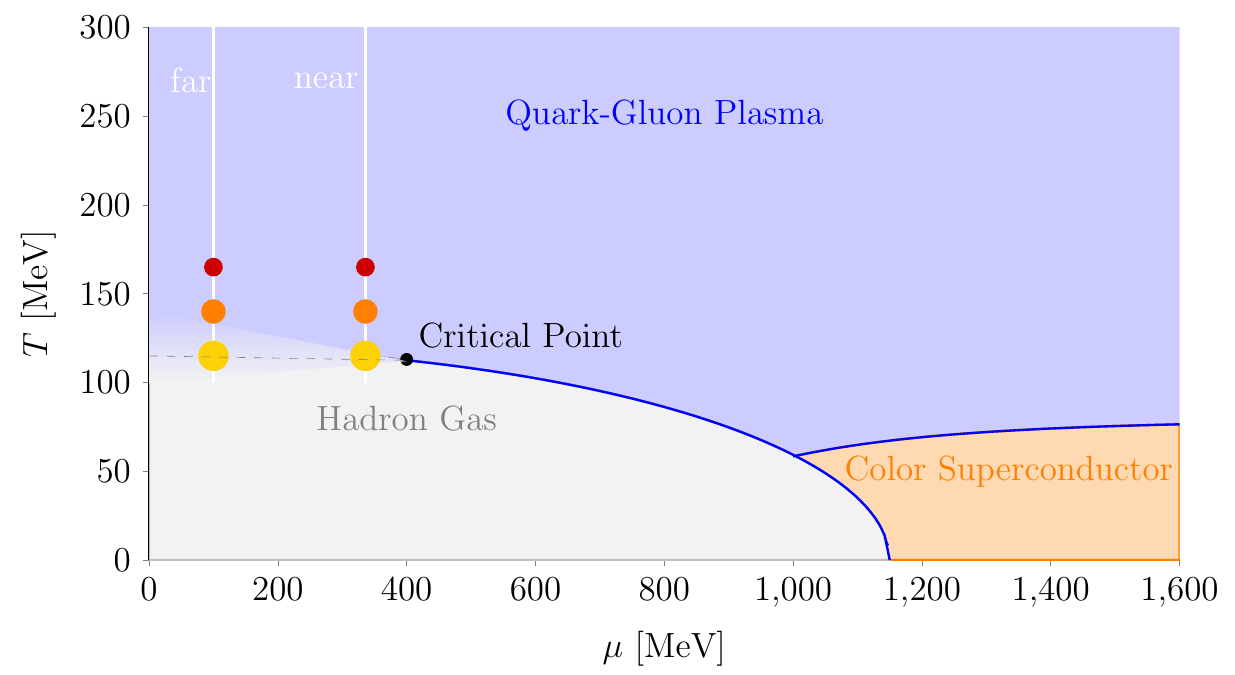}
		\caption{Schematic illustration of the background trajectories used in this work in the $(T,\mu)$ plane of QCD phase diagram. We consider constant-$\mu$ trajectories (vertical lines) with a prescribed cooling history $T(t)$. The sequence of colored markers indicates representative times along each trajectory (earlier $\to$ later as $T$ decreases). The progressive decreasing color intensity of the marker color encodes the decrease of temperature, while the increasing marker size is a purely schematic indication of the growth of an effective macroscopic domain size as the system evolves. The lightly shaded band is a schematic indication of the crossover region, emphasizing that, away from the critical point, the change from quark-gluon
			plasma to hadron gas is smooth rather than a sharp phase transition. The figure is intended as a visual guide to the setup; the correlator evolution is performed long these trajectories using the time-dependent inputs specified in Sec.~\ref{sec:EoS-transport} rather than by solving an explicitly expanding hydrodynamic background.}
		\label{fig:phase_diagram}
	\end{figure}

\begin{figure}[t]
		\centering
		\includegraphics[width=.48\textwidth]{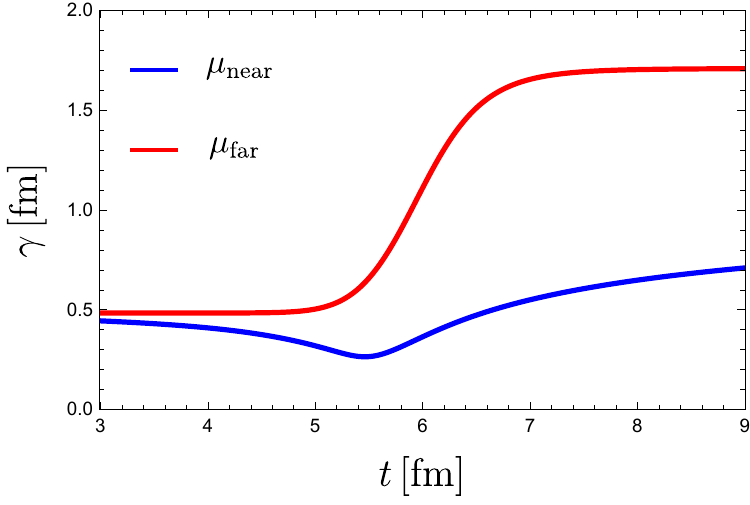}\,\,\,\,\includegraphics[width=.48\textwidth]{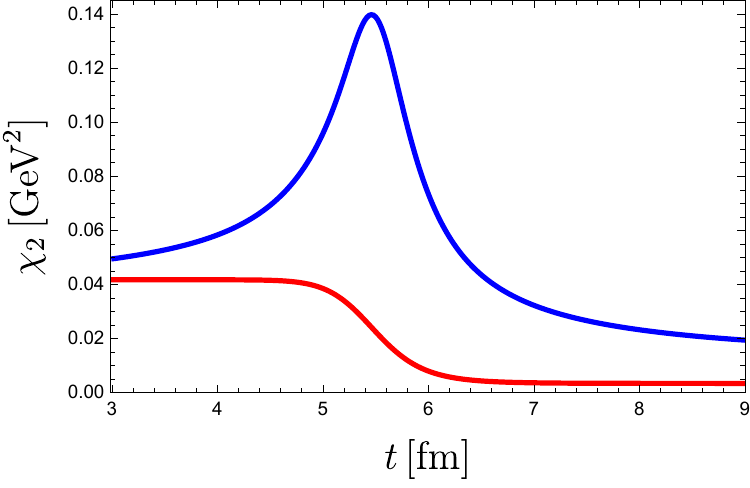}
		\includegraphics[width=.48\textwidth]{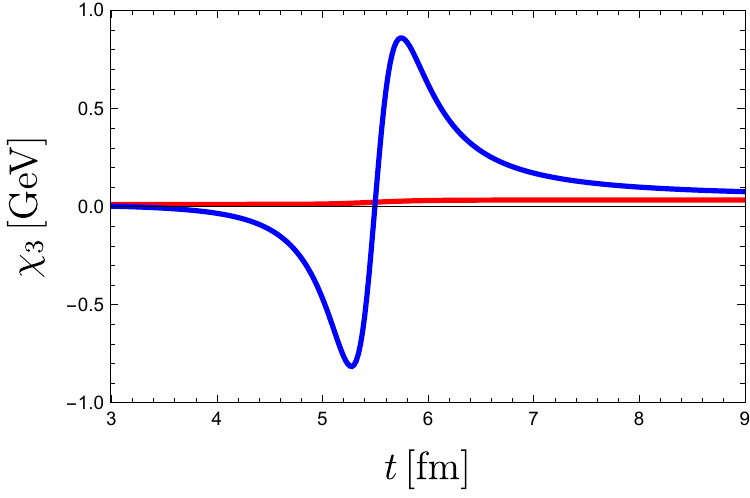}\,\,\,\,\includegraphics[width=.48\textwidth]{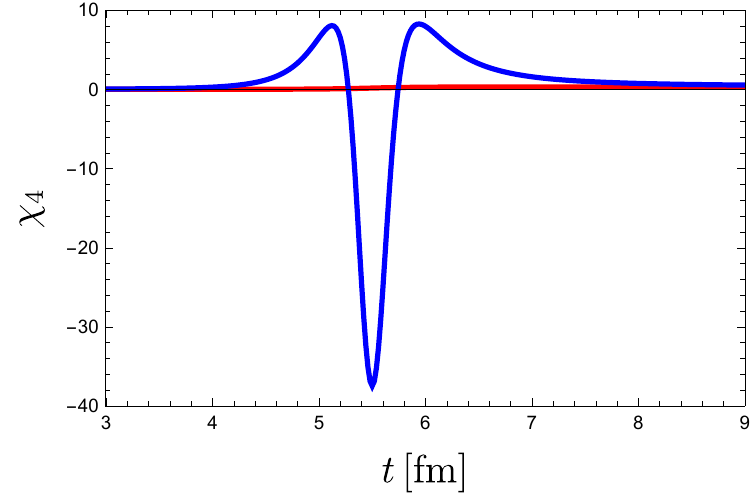}
		\caption{Time-dependent thermodynamic and transport inputs along the background evolution at fixed baryon chemical potential. Shown are the susceptibilities $\chi_2(t)$, $\chi_3(t)$,
			$\chi_4(t)$ (from the 3D Ising mapping evaluated along fixed $\mu$) and the diffusion coefficient
			$\gamma(t)$ used in the evolution equations of stochastic MC diffusion.
			Results are displayed for two representative constant-$\mu$ trajectories, $\mu=\mu_{\text{far}}\equiv0.100$~GeV
			(red) and $\mu=\mu_{\text{near}}\equiv0.366$~GeV (blue). The pronounced non-monotonic structures in $\chi_n(t)$ for the
			larger $\mu$ reflect passing close to the critical region (with $\mu_c= 0.4$~GeV in our
			parameter choice) and induce a corresponding time dependence of $\gamma(t)$ through the adopted
			parametrization.
		}	\label{fig:coeff}
	\end{figure}

	To make the discussion concrete while keeping the background dynamics under control, we work with representative trajectories through the QCD phase diagram at fixed baryon chemical potential. We prescribe a time-dependent temperature history $T=T(t)$ and keep chemical potential $\mu$ constant, so that the background follows a vertical line in the $(T,\mu)$ plane. We consider two values, $\mu=\mu_{\rm far}$ (far from the critical region) and $\mu=\mu_{\rm near}$ (passing close to it). This provides a simple way to separate noncritical finite-lifetime effects from those enhanced by critical slowing down (Fig.~\ref{fig:phase_diagram}). Along each trajectory, the susceptibilities $\chi_n(t)$ and diffusion coefficient $\gamma(t)$ (Fig.~\ref{fig:coeff}) are evaluated from our chosen equation of state via Ising mapping and serve as time-dependent inputs for the correlator evolution.

	Furthermore, to explore the phenomenological consequence of fluctuation dynamics in MC theory, we model the QGP evolution through the QCD phase diagram using the chemical potential as a control parameter. This mimics the beam energy scan in heavy-ion collisions, where different collision energies correspond to different trajectories in the $(T,\mu)$ plane. In particular, we consider a set of trajectories with fixed chemical potential $\mu$, while the bulk evolution along each trajectory is characterized by a decreasing temperature $T(t)$. When the temperature reaches the freeze-out value $T_f$, the dynamical correlators $W_N$ are converted into acceptance-dependent cumulants $C_N(\Delta)$ via Eq.~\eqref{eq:C_N-W_N}.

	With the above setup, we now illustrate the parameterization of thermodynamic and transport quantities.

	In the MC diffusion equations, the linear relaxation rate of a density mode is governed by diffusion coefficient $\gamma=\lambda\alpha'$. 
	The nonlinear transport couplings appearing in the source terms are encoded in thermodynamic derivatives of $\alpha(n)=\mu(n)/T$ with respect to $n$, which can be expressed in terms of baryon-number susceptibility $\chi_k\equiv(\frac{\partial^{k-1} n}{\partial\mu^{k-1}})_T$ as
	\begin{align}\label{eq:alpha-chi}
		\alpha'= \frac{1}{T\chi_2}\,,\qquad	\alpha''= -\frac{1}{T} \frac{\chi_3}{\chi^3_2}\,,\qquad
		\alpha''' = \frac{1}{T} \left(3 \frac{\chi^2_3}{\chi^5_2} - \frac{\chi_4}{\chi^4_2}\right).
	\end{align}
	In this work, thermal conductivity is parameterized as $\lambda = d_c T(t_0)\chi^{1/2}_2(t_0)$, where $d_c=0.5$ controls the overall magnitude of diffusion. The diffusion coefficient $\gamma$ and its derivatives are then given by $\gamma = \lambda \alpha',\, \gamma' = \lambda \alpha'', \gamma'' = \lambda \alpha'''$ for constant $\lambda$.
	
	The simulation requires the time dependence of $\gamma(t)$ and $\chi_k(t)$ along a background trajectory.
	We obtain baryon susceptibilities from a standard mapping to the 3D Ising model.
	Following Refs.~\cite{Sakaida:2017rtj,Pihan:2022xcl}, we decompose the susceptibilities as follows
	\begin{equation}\label{eq:chi_decompose}
		\chi_k = \chi_k^{\mathrm{cri}} + \chi_k^{\mathrm{reg}} .
	\end{equation}
	\paragraph{Regular part.} The regular susceptibilities are interpolated between a hadronic baseline and a QGP baseline, i.e.,
	\begin{equation}\label{eq:chi^reg}
		\chi_k^{\mathrm{reg}}(T)=\chi_k^{\mathrm{H}}+\frac{1}{2}\Big(\chi_k^{\mathrm{QGP}}-\chi_k^{\mathrm{H}}\Big)\Big[1+\tanh\!\Big(\frac{T-T_c}{\Delta T_{\rm tr}}\Big)\Big]\,,
	\end{equation}
	with $\Delta T_{\rm tr}=0.01~\mathrm{GeV}$ controlling the width of the transition region between the two phases~\cite{Sakaida:2017rtj}. For later convenience, we normalize the susceptibilities $\chi_k$ in terms of the dimensionless lattice results $\chi_k^{\rm latt}$ for hadronic and QGP phase and $k=2,3,4$, respectively:
	\begin{align}\label{eq:chi-chi_latt} 
		\chi_2^{\mathrm{H}}&=T_{\mathrm{H}}^{2}\chi_2^{\mathrm{H,latt}},\qquad
		\chi_2^{\mathrm{QGP}}=T_{\mathrm{QGP}}^{2}\chi_2^{\mathrm{QGP,latt}},\nonumber\\
		\chi_3^{\mathrm{H}}&=T_{\mathrm{H}}\chi_3^{\mathrm{H,latt}},\qquad
		\chi_3^{\mathrm{QGP}}=T_{\mathrm{QGP}}\chi_3^{\mathrm{QGP,latt}},\nonumber\\
		\chi_4^{\mathrm{H}}&=\chi_4^{\mathrm{H,latt}},\qquad\,\,\,\,\,\,\,
		\chi_4^{\mathrm{QGP}}=\chi_4^{\mathrm{QGP,latt}}\,,
	\end{align}
	where the typical temperatures of the hadronic and QGP phases are given by $T_{\mathrm{H}}=0.10~\mathrm{GeV}$ and $T_{\mathrm{QGP}}=0.25~\mathrm{GeV}$, respectively. The choice of lattice susceptibilities is as following: we take $\chi_2^{\rm H,latt}=1/3$, $\chi_2^{\rm QGP,latt}=2/3$ from Ref.~\cite{Ling:2013ksb}. For the fourth-order susceptibility, we use $\chi_4^{\mathrm{H,latt}}=\chi_2^{\mathrm{H,latt}}$ and $\chi_4^{\mathrm{QGP,latt}}=\frac{2}{3\pi^2}\chi_2^{\mathrm{QGP,latt}}$~\cite{Cheng:2008zh,Bazavov:2017dus}. Since Lattice QCD does not provide the third-order baryon number susceptibility,  we estimate it as $\chi^{\mathrm{latt}}_3\sim\chi^{\mathrm{latt}}_4$ ~\cite{Mukherjee:2016nhb} for both phases.
	
	\paragraph{Critical part.}
	There are various choices for mapping the critical part of the equation of state~\cite{Parotto:2018pwx,An:2025kaw}. In this work, we map the critical contribution to QCD susceptibilities, $\chi^{\rm cri}_k$, from the 3D Ising susceptibilities $\kappa_k$ via the following parameterization:
	\begin{equation}\label{eq:chi_k^cri}
		\chi_k^{\mathrm{cri}}(T,\mu)= T_A^{4-k}\,\kappa_k(R,\theta)\,,
	\end{equation}
	where $T_A=0.3~\mathrm{GeV}$ is introduced for dimensional consistency. The parametrization variables $(R,\theta)$ are related to Ising variables $(r,h)$ by
	\begin{equation}\label{eq:rh_para}
		r(R,\theta)=R(1-\theta^2)\,,\qquad
		h(R,\theta)=R^{5/2}(3\theta-2\theta^3)\,,
	\end{equation}
	where $r$ is the reduced Ising temperature while $h$ is the magnetic field. The critical region $(\Delta h, \Delta r)$ in the Ising phase diagram is further mapped to the critical region $(\Delta T,\Delta\mu)$ in the QCD phase diagram via
	\begin{equation}\label{eq:mapping}
		\frac{T-T_c}{\Delta T}=\frac{h}{\Delta h}\,,
		\qquad
		\frac{\mu-\mu_c}{\Delta\mu}=-\frac{r}{\Delta r}\,.
	\end{equation}
	The Ising susceptibilities $\kappa_{k=2,3,4}$ are parametrized as~\cite{Schofield:1969zza,
		Josephson_1969}
	\begin{align}\label{eq:kappa234}
		\kappa_2(R,\theta) &= \frac{M_0}{H_0}\,\frac{1}{R^{4/3}\,(3+2\theta^2)}\,,\nonumber\\
		\kappa_3(R,\theta) &= -\frac{M_0}{H_0^2}\,
		\frac{4\theta(9+\theta^2)}{R^{3}(3-\theta^2)(3+2\theta^2)^3}\,,\nonumber\\
		\kappa_4(R,\theta) &= -\frac{12M_0}{H_0^3}\,
		\frac{81-783\theta^2+105\theta^4-5\theta^6+2\theta^8}{R^{14/3}(3-\theta^2)^3(3+2\theta^2)^5}\,,
	\end{align}
	where we take the normalization constants $M_0\simeq 0.605$ and $H_0\simeq 0.394$~\cite{Parotto:2018pwx}. In this parameterization of the equation of state mapped from the Ising phase diagram, the location of the critical point is treated as a free parameter. We adopt  $T_c=0.12~\mathrm{GeV}$ and $\mu_c=0.40~\mathrm{GeV}$, within the theoretically predicted range of possible critical point locations $(T_c, \mu_c)$ compiled in Ref.~\cite{Zhang:2026dny}. The other non-universal parameters are chosen as~\cite{Mukherjee:2015swa,Wu:2018twy,Wu:2024ixv}
	\begin{equation}\label{eq:mapping_values}
		\Delta T=\frac{1}{8}T_c=0.015~\mathrm{GeV}\,,\qquad
		\Delta \mu=0.10~\mathrm{GeV}\,,\qquad
		\Delta r=\left(\frac{5}{3}\right)^{3/4}\,,\qquad
		\Delta h=1\,.
	\end{equation}

	\paragraph{Background trajectory and time variable.}
	For illustration we prescribe a phenomenological cooling profile inspired by Hubble/Bjorken-like expansion, with temperature 
	\begin{equation}\label{eq:Hubble_expansion}
		T(t)=T_0\left(\frac{t}{t_0}\right)^{-3c_s^2}\,,\qquad c_s^2=\frac13\,,
	\end{equation}
	where $t$ is the evolution time variable used in our correlator equations, $c_s$ is the speed of sound that controls the cooling rate.  In our setup, the initial conditions of the prescribed evolution are specified by $T_0=0.22~\mathrm{GeV},\, t_0=3~\mathrm{fm}$, while the freeze-out temperature is chosen as $T_f = 0.11~\mathrm{GeV}$. We use Eq.~\eqref{eq:Hubble_expansion} as a phenomenological parametrization of the cooling history experienced by a representative fluid element.
	Accordingly, in the dynamical evolution we do not include the terms controlled by the background expansion rate $\Theta$ (in an expanding hydrodynamic background, $\Theta\equiv\nabla_\mu u^\mu$; e.g., Bjorken/Milne terms $\sim 1/\uptau$) that arise in an explicitly expanding formulation. This setup should therefore be viewed as an effective modeling for the time dependence of the thermodynamic inputs $\chi_n(t)$ and transport coefficients (such as $\gamma(t)$) along the prescribed trajectory, rather than as a detailed spacetime description of the full collision.  A rough consistency criterion for neglecting explicit expansion terms is that the intrinsic relaxation rates of the retained modes, such as $\gamma(t)q^2$ (and $1/\tau$ in MC theory), are not parametrically smaller than the background expansion scale $\Theta$ over the time interval of interest (see more discussions in Sec.~\ref{sec:numerics}).

	In the present work we deliberately simulate the fluctuation dynamics along constant-$\mu$ background trajectories: we specify $T(t)$ via Eq.~\eqref{eq:Hubble_expansion} and keep the baryon chemical potential $\mu$ fixed, so that the background follows a vertical line in the $(T,\mu)$ plane. To isolate finite current-memory effects in a controlled way, we perform representative runs at two chemical potentials, $\mu=\mu_{\rm far}$ (far from the critical region) and $\mu=\mu_{\rm near}$ (passing close to it), as illustrated in Fig.~\ref{fig:phase_diagram}. For each $\mu$ we consider two values of the MC relaxation time, $\tau=\tau_{\rm small}$ and $\tau=\tau_{\rm large}$. For every choice $(\mu,\tau)$, we compare the MC evolution with the purely diffusion baseline obtained by setting $\tau=0$ while keeping all other inputs identical. 
	
	\paragraph{Equilibrium correlators and observables.}
	The equilibrium correlators provide a convenient reference and also determine our initial
	conditions chosen to be in equilibrium. The equilibrium connected equal-time correlators are
	\begin{align}\label{eq:W234}
		W^{\mathrm{eq}}_2 &= \frac{1}{\alpha'}=T\chi_2\,,\nonumber\\
		W^{\mathrm{eq}}_3 &= -\frac{\alpha''}{(\alpha')^{3}}=T^{2}\chi_3\,,\nonumber\\
		W^{\mathrm{eq}}_4 &= \frac{3(\alpha'')^{2}-\alpha'\alpha'''}{(\alpha')^{5}}=T^{3}\chi_4\,,
	\end{align}
	where the last equality in each equation follows after substituting $\alpha',\alpha'',\alpha'''$ in terms of $\chi_k$ using Eq.~\eqref{eq:alpha-chi}.
	
	Furthermore, what is commonly reported in experiment is the ratios of cumulants
	\begin{equation}\label{eq:cumulat-ratios}
		S\sigma \equiv \frac{C_3}{C_2}\,,
		\qquad
		\kappa\sigma^2 \equiv \frac{C_4}{C_2}\,.
	\end{equation}
	Their equilibrium values are obtained by replacing $W_N(q_1,\cdots,q_N;t)$
	with the equilibrium correlators $W_N^{\rm eq}$ in Eq.~\eqref{eq:C_N-W_N}, and
	reduce to the susceptibility ratios $\chi_3/\chi_2$ and $\chi_4/\chi_2$,
	respectively.

	\section{Numerical setup and results}
	\label{sec:numerics}
	In this section we solve the evolution equations for the equal-time Wigner functions in MC theory including its Fickian diffusive limit, and quantify the resulting modifications of freezeout correlators and acceptance-dependent cumulants. All simulations are performed along the prescribed constant-$\mu$ background trajectories introduced in Sec.~\ref{sec:EoS-transport}. By comparing a low-$\mu$ far-from-critical trajectory, $\mu=\mu_{\rm far}$, with a near-critical trajectory, $\mu=\mu_{\rm near}$, we disentangle generic finite-current-relaxation effects from the additional modifications induced near the critical region.

    \paragraph{Physical momentum window.}
	Our effective description is intended for wavelengths that are (i) supported by the finite size of the QGP droplet and (ii) not shorter than the the length scale below which hydrodynamics ceases to apply. We therefore restrict the modes to a momentum window
\begin{equation}
	\label{eq:qwindow_def}
	q \in [q_{\min},q_{\max}]\,, \qquad
	q_{\min}\sim \mathcal{O}\!\left(\frac{1}{L}\right)\,, \qquad
	q_{\max}\sim \mathcal{O}\!\left(\frac{1}{\ell_{\rm cell}}\right)\,,
\end{equation}
where $L$ denotes a characteristic macroscopic system size and $\ell_{\rm cell}$ denotes a UV cutoff length scale controlling the breakdown of the coarse-grained description. In practice, we take
\begin{equation}
	q_{\min}\simeq 0\,, \qquad
	q_{\max}\simeq 2~{\rm fm}^{-1}\,.
	\label{eq:qwindow_vals}
\end{equation}
while the UV cutoff corresponds to $\ell_{\rm cell}\sim\mathcal{O}(0.5~{\rm fm})$, the finite-size infrared scale $O(1/L)$ is neglected in the large-system continuum approximation, consistent with the homogeneous-background assumption underlying Eq.~\eqref{eq:trans_inv}.

	\paragraph{Consistency estimate: expansion vs relaxation.}
	As discussed in Sec.~\ref{sec:EoS-transport}, a rough consistency criterion for neglecting explicit expansion terms is that the intrinsic relaxation rates of most retained modes, $\Gamma(t)=\gamma(t)q^2$, are not parametrically smaller than a typical expansion scale, $\Theta$, over the time interval of interest. For orientation, one may take a Bjorken-like estimate $\Theta\sim 1/\uptau$ at midrapidity, which in our time window corresponds to $\Theta\sim\mathcal{O}(0.1$--$0.3)\,{\rm fm}^{-1}$. The retained modes for which expansion effects may be relevant can be estimated from the competition between the expansion and diffusion rates, yielding $q\lesssim\sqrt{\Theta/\gamma(t)}\sim\mathcal{O}(0.5~{\rm fm}^{-1})$, where the minimum value of the time-dependent diffusion coefficient $\gamma(t)$ is around $0.5~{\rm fm}$ (see Fig.~\ref{fig:coeff}). 
    
    Consequently, expansion effects, if included, are expected to be most relevant in the small-$q$ region of the retained momentum range. We neglect these explicit expansion corrections in the present work and regard their quantitative impact on the softest modes as a limitation of the simplified background treatment.

	\paragraph{Choice of relaxation times.}
	The MC characteristic momentum scale $q_\star$ indicates where finite current relaxation competes with diffusion.
	Since $\gamma(t)$ varies along the trajectory, $q_\star$ is time dependent; for orientation, we estimate it using a
	representative value of $\gamma$, for example near the region of strongest variation or at freezeout.
	Throughout this work we use two relaxation times, $\tau=\tau_{\rm small}$ and $\tau=\tau_{\rm large}$, chosen such
	that $q_\star$ lies inside $[q_{\min},q_{\max}]$ for one choice and outside for the other. This provides a  a clean test of
	when memory effects should be visible in the physical momentum window.
	
	With these conventions established, we now turn to the numerical evolution of $W_N(t,q)$.
	
	\subsection{Two-point functions}
	\label{subsec:evo_W2}
	
	We begin with the two-point Wigner function $W_2(q,t)$, which controls the variance of the conserved charge in a finite acceptance window. Along each prescribed trajectory we compare the Fickian diffusive evolution ($\tau=0$) with the MC evolution at finite current-relaxation time ($\tau\neq0$). Unless stated otherwise, we present the normalized Wigner function
	\begin{equation}\label{eq:tildeW2}
		\widetilde W_2(q,t)\equiv \frac{W_2(q,t)}{W_2(q,t_0)}\,,   
	\end{equation}
	and use the instantaneous local-equilibrium value $W_2^{\rm eq}(t)$, determined by the time-dependent susceptibility along the same background trajectory, as a reference benchmark. 
	
\begin{figure}[!tb]
		\centering
		\includegraphics[width=1.04\textwidth]{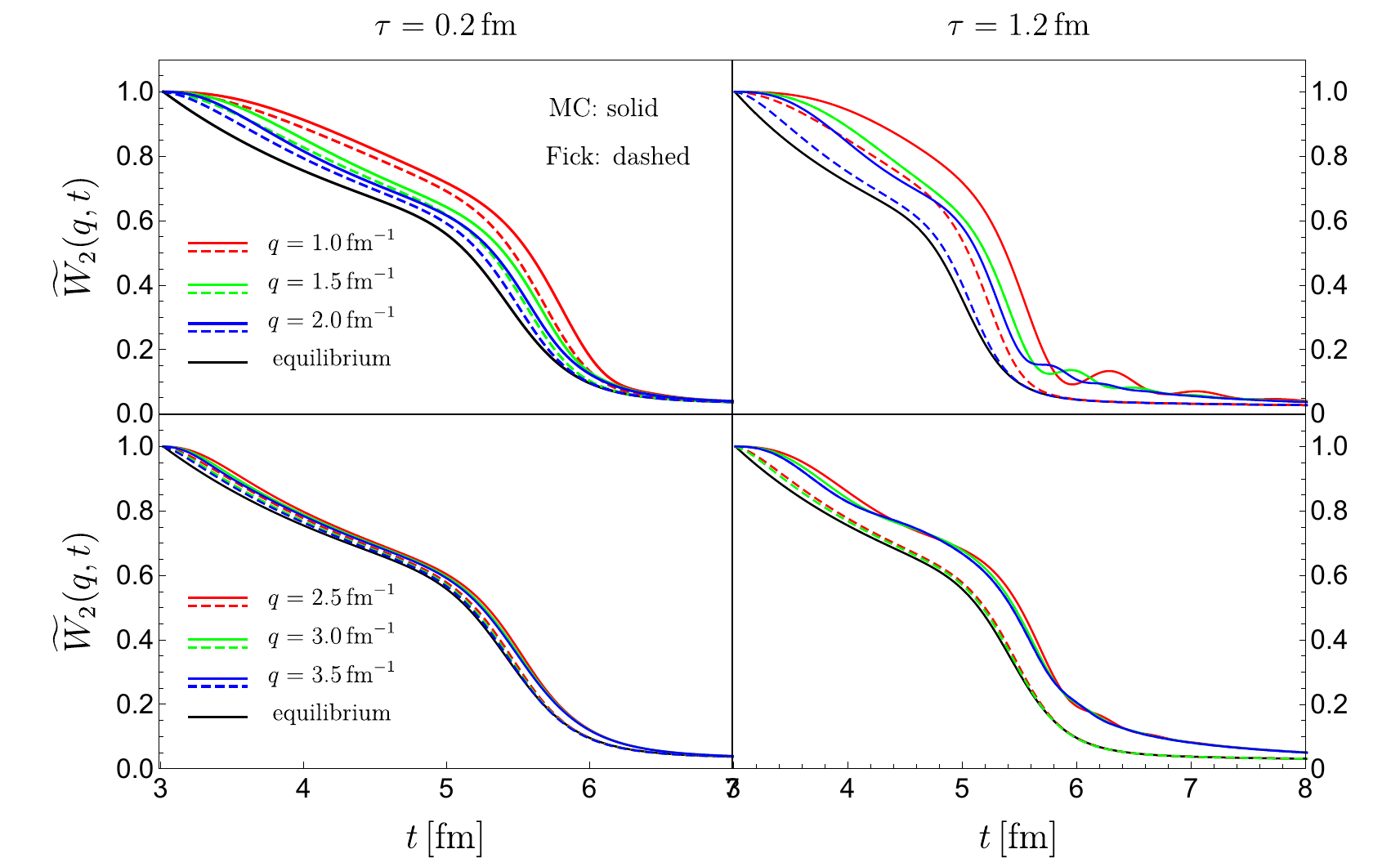}
		\caption{Time evolution of the normalized two-point Wigner function \(\widetilde W_2(q,t)\) along the constant-\(\mu\) trajectory at \(\mu=\mu_{\rm far}=0.100\,{\rm GeV}\), far from the critical region. Colored solid curves show the MC evolution with finite current-relaxation time, while colored dashed curves show the Fickian diffusion limit, \(\tau=0\), for the same values of \(q\). The black curve denotes the instantaneous local-equilibrium value \(W_2^{\rm eq}(t)\), determined by the time-dependent susceptibility along the same background trajectory; it serves as a common reference for both the MC and Fickian evolutions. The left and right columns correspond to \(\tau=0.2\,{\rm fm}\) and \(\tau=1.2\,{\rm fm}\), respectively. The upper panels show representative modes in the physical window \(q\in[q_{\min},q_{\max}]\), while the lower panels show larger-\(q\) modes, included only as a diagnostic.} 
		\label{fig:W_2_far}
	\end{figure}

\begin{figure}[!tb]
		\centering
		\includegraphics[width=1.04\textwidth]{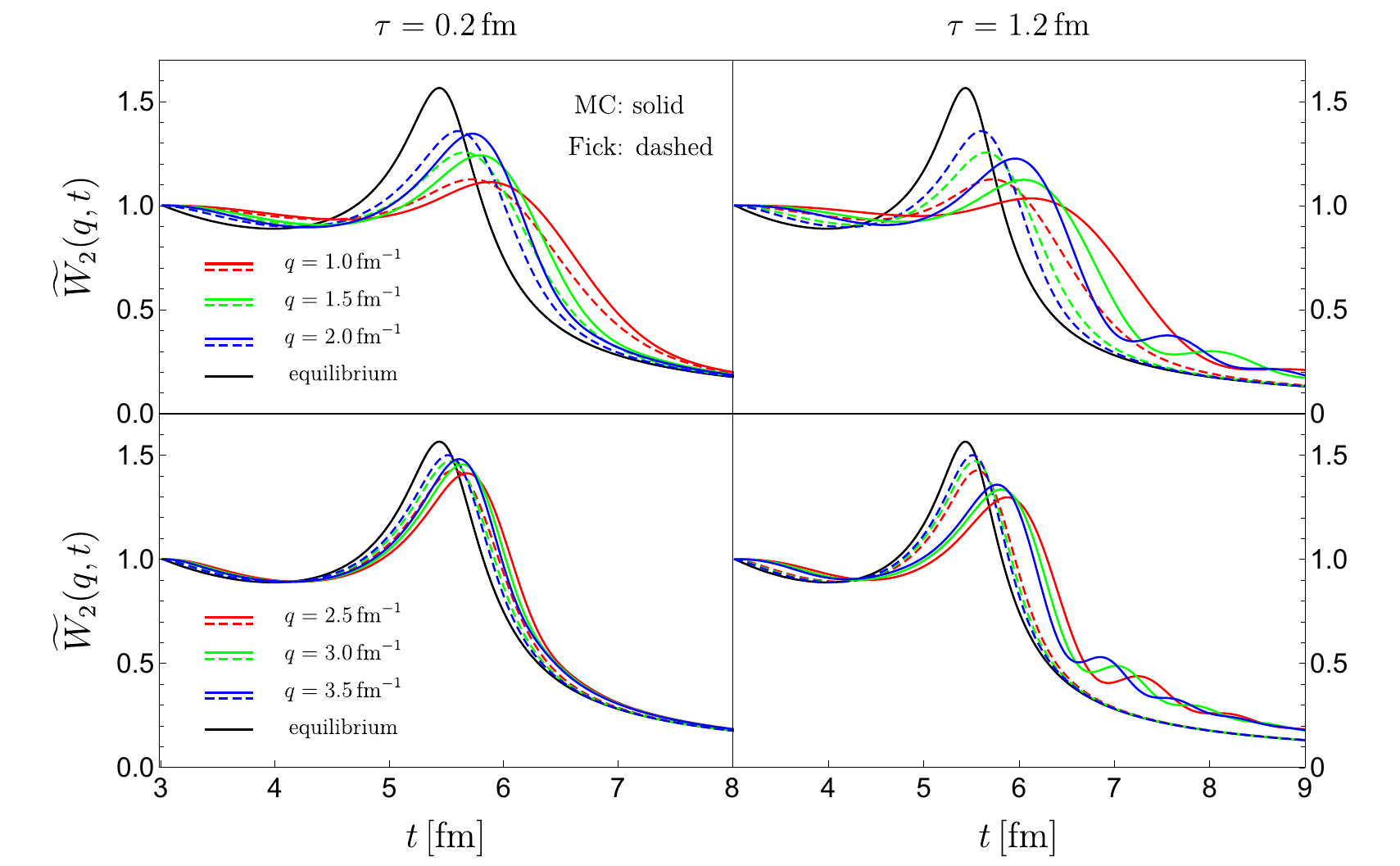}
		\caption{Same as Fig.~\ref{fig:W_2_far}, but for the constant-\(\mu\) trajectory at \(\mu=\mu_{\rm near}=0.366\,{\rm GeV}\), which passes close to the critical region. Compared with Fig.~\ref{fig:W_2_far}, the near-critical trajectory makes the finite-$\tau$ effects more visible, especially for $\tau=\tau_{\rm large}=1.2\,{\rm fm}$, where the MC curves show both an additional delay and a mild reshaping of the relaxation profile relative to the Fickian baseline.} 
		\label{fig:W_2_near}
	\end{figure}

	In the multi-panel plots of Figures~\ref{fig:W_2_far} and \ref{fig:W_2_near}, the top panels display representative wave numbers inside the window $[q_{\text{min}},q_{\text{max}}]$, which are the modes entering our acceptance-dependent cumulants. The bottom panels show $q$ values outside the window only as a \emph{diagnostic} and they are not used in the phenomenological integrals in our evaluation of acceptance-dependent cumulants. The
	finite-$\tau$ effect should be read relative to the Fickian diffusion
	baseline.

	Let us first discuss the in-window modes shown in the top panels of Figs.~\ref{fig:W_2_far} and \ref{fig:W_2_near}. For the smaller relaxation time, $\tau=\tau_{\rm small}$, the corresponding $q_\star$ lies outside the physical window. The MC curves remain close to the corresponding diffusion curves both far from and near the critical region and  the main visible effect is only a modest additional delay.  This indicates that, for this choice of $\tau$, current memory is a small correction in the physical momentum window. 
	
	For the larger relaxation time, $\tau=\tau_{\rm large}$, the scale $q_\star$ falls inside the physical window, so memory effect becomes clearly visible.  Along the far-from-critical trajectory in Fig.~\ref{fig:W_2_far}, where the equilibrium target varies smoothly, finite current relaxation mainly shifts the response to later times relative to the diffusion baseline, with only a mild distortion of the relaxation tail.  Along the near-critical trajectory in Fig.~\ref{fig:W_2_near}, the equilibrium target has a pronounced non-monotonic structure. The same delayed current response therefore does more than shift the curves; it also reshapes the peak region and the subsequent relaxation. The approach to the evolving equilibrium target becomes weakly oscillatory, corresponding to an underdamped response.

	More precisely, for fixed transport coefficients, the density-current modes become underdamped when $q>q_\star=1/\sqrt{4\tau\,\gamma}$ (see Eqs.~\eqref{eq:powercounting_qstar} and~\eqref{eq:omega_pm}). Since $\gamma=\gamma(t)$ varies along the background
	trajectories considered here, this condition should be understood locally in
	time. The separation between the overdamped and
	underdamped regimes was analyzed systematically in Ref.~\cite{Abbasi:2025teu}; here we focus
	on how this crossover appears in the evolution of the equal-time correlators.

	The lower panels in Figs.~\ref{fig:W_2_far} and \ref{fig:W_2_near} are shown only as a diagnostic of the large-$q$ behavior outside the physical momentum window.  In the Fickian baseline, increasing $q$ simply increases the relaxation rate, $\Gamma_q=\gamma q^2$, so the large-$q$ modes track the instantaneous equilibrium target more efficiently. Thus the dashed curves in the bottom panels lie very close to the black curve~\cite{An:2020vri}.
	
	The MC evolution with finite $\tau$, however, behaves differently.  Because the diffusive current has its own finite relaxation time, the large-$q$ modes do not necessarily approach the instantaneous Fickian limit faster and faster (see more discussions on large $q$ behavior in Sec.~\ref{subsec:q-dependence}).  When the diffusive relaxation rate $\gamma q^2$ becomes comparable to the current-relaxation rate $1/\tau$, the evolution is no longer governed only by the diffusive rate.  This results in an additional delay beyond Fickian baseline and, for the larger relaxation time, a weakly underdamped relaxation pattern. This effect is mild along the far-from-critical trajectory, where the equilibrium target varies smoothly, but becomes visible
	near the critical trajectory, where the equilibrium benchmark changes
	non-monotonically in time.

	\subsection{Three- and four-point functions}
	\label{subsec:evo_W34}
	Having established the evolution analysis for the two-point correlators $W_2$, we now turn to the higher-point connected correlators $W_3$ and $W_4$. Since equilibrium higher-point correlators are known to be more sensitive to criticality than $W_2$, it might be interesting to explore how the dynamic effects, especially the finite current relaxation, may reveal or reshuffle this sensitivity.

	\begin{figure}[!tb]
		\centering
		\includegraphics[width=1\textwidth]{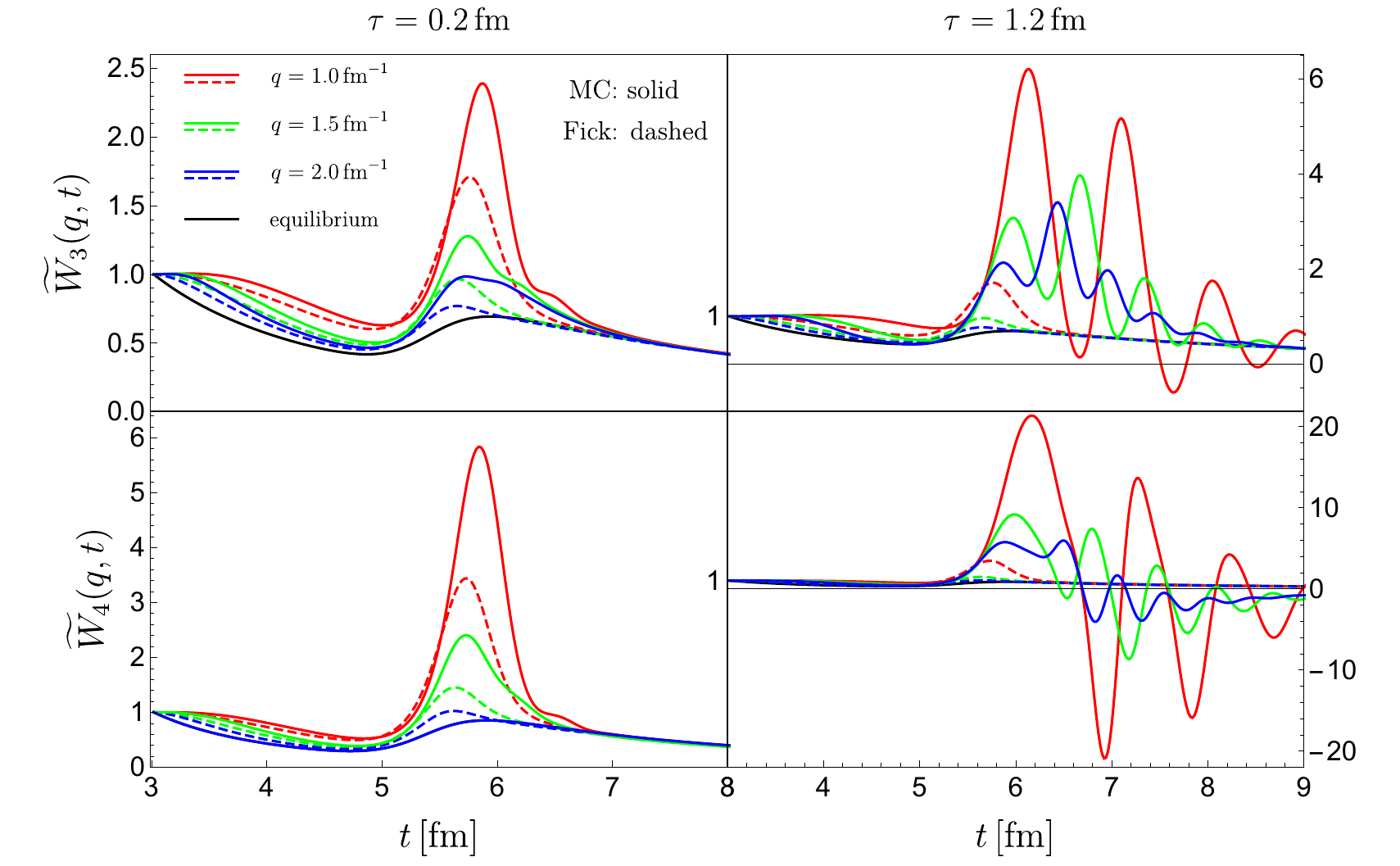}
		\caption{Time evolution of the normalized higher connected correlators along the
			far-from-critical trajectory, $\mu=\mu_{\rm far}\equiv 0.100~{\rm GeV}$, for representative in-window wave numbers
			$q=1.0,\ 1.5,\ 2.0~{\rm fm}^{-1}$. Top panels:
			$\widetilde W_3(q,t)$; bottom panels: $\widetilde W_4(q,t)$. Dashed curves show the Fickian diffusion baseline ($\tau=0$), while solid curves show MC evolution with finite current-relaxation time. Left column:
			$\tau=\tau_{\rm small}$; right column: $\tau=\tau_{\rm large}$. The black curve in each panel denotes the corresponding instantaneous local-equilibrium benchmark along	the same trajectory. Finite relaxation produces both a delayed response and a $q$-dependent reshaping of the higher connected correlators relative to the diffusion baseline. The large apparent amplitude of the oscillations in the right panels is enhanced by the
			normalization to the small initial values of the higher-point correlators along the far-from-critical trajectory; these panels should therefore be interpreted mainly as showing the delayed and underdamped response induced by finite current relaxation.}
		\label{fig:W_34_far}
	\end{figure}
	
\begin{figure}[t]
		\centering
		\includegraphics[width=1\textwidth]{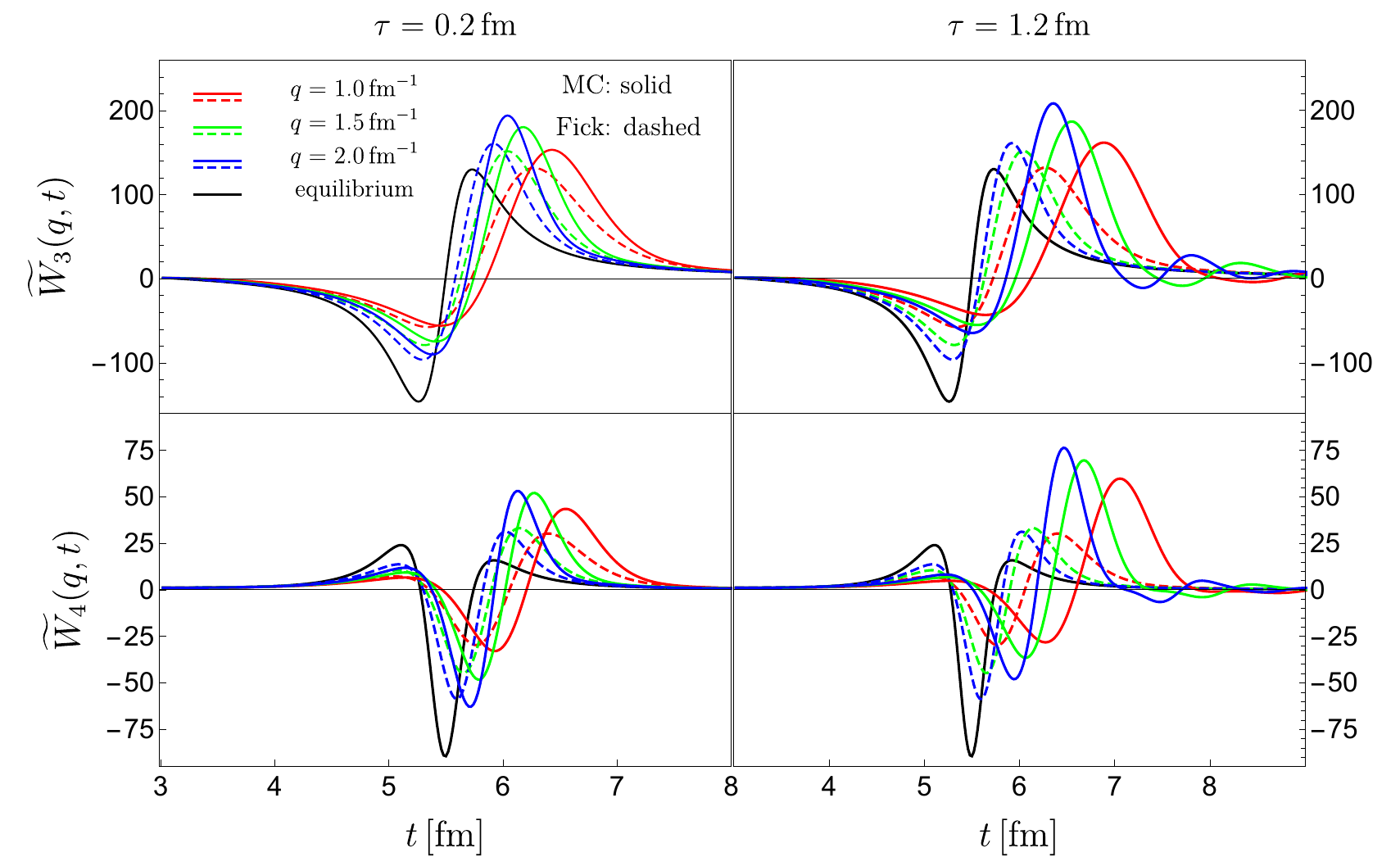}
		\caption{Same as Fig.~\ref{fig:W_34_far}, but for the near-critical trajectory $\mu=\mu_{\rm near}\equiv 0.366~{\rm GeV}$, which passes close to the critical region. Compared with the far-from-critical case, finite current relaxation produces more visible deviations from the Fickian baseline; the extrema are shifted, the peak structure is reshaped, and the late-time relaxation is more strongly modified, especially for $\tau=\tau_{\rm large}$.}
		\label{fig:W_34_near}
	\end{figure}

	Similar to Eq.~\eqref{eq:tildeW2}, we define the normalized three- and four-point Wigner functions
	\begin{equation}\label{eq:tildeW34}
		\widetilde W_3(q,t)\equiv \frac{W_3(q,t)}{W_3(q,t_0)} \qquad\text{and}\qquad \widetilde W_4(q,t)\equiv \frac{W_4(q,t)}{W_4(q,t_0)}  \,.
	\end{equation}
	Their time evolutions are shown in Figs.~\ref{fig:W_34_far} and \ref{fig:W_34_near} for $\mu_{\rm far}$ and $\mu_{\rm near}$, respectively. We focus on the comparison between the solid MC curves and the dashed Fickian curves at the same values of $q$, yet only within the physical momentum window. As one can expect from the previous results shown in Figs.~\ref{fig:W_2_far} and \ref{fig:W_2_near} for two-point functions, the Fickian evolution does not follow the instantaneous local-equilibrium benchmark exactly, because the equilibrium benchmark itself changes with time, and the finite current relaxation in addition modifies the results on top of the Fickian nonequilibrium evolution.
	
	Along the far-from-critical trajectory in Fig.~\ref{fig:W_34_far}, a finite relaxation time shifts this response and changes the height and location of the extrema.  For $\tau=\tau_{\rm small}$ the modification is mostly a smooth deformation of the Fickian result, while for $\tau=\tau_{\rm large}$ the solid curves depart much more strongly from the dashed curves and develop an underdamped relaxation pattern.  The large apparent amplitude of these oscillations should be interpreted with care, since it is enhanced by the normalization to the small initial values $W_N(q,t_0)$ of the higher-point correlators along the far-from-critical trajectory.  The robust information is therefore the finite-$\tau$ shift, the $q$-dependent reshaping relative to the Fickian baseline, and the onset of an underdamped response.

	The near-critical trajectory in Fig.~\ref{fig:W_34_near} shows the same comparison in a more structured background.  Here the Fickian curves already display a pronounced lag and distortion relative to the instantaneous equilibrium benchmark, reflecting the rapid non-monotonic evolution of the equilibrium higher-point correlators.  Finite current relaxation further modifies this Fickian response; the extrema are shifted, their magnitudes are changed, and the subsequent relaxation is reshaped, especially for $\tau=\tau_{\rm large}$.  Thus, the effect of current memory is not simply that the correlators fail to follow their equilibrium values, which already happens in Fickian diffusion. Rather, the finite $\tau$ accounts for the nontrivial deviation of the correlators form their Fickian baseline.

	Taken together, Figs.~\ref{fig:W_34_far} and ~\ref{fig:W_34_near} show that the higher-point correlators provide a sharper diagnostic of finite current relaxation than $W_2$.
	This suggests that the corresponding freezeout cumulants, $C_3$ and $C_4$, should be especially sensitive to current memory,  especially near the critical region. 	Before converting the evolved correlators into acceptance-dependent cumulants,
	we first examine their \(q\)-dependence at freezeout.

	\subsection{Spectrum of correlators at freezeout}
	\label{subsec:q-dependence}
	In the previous subsections we studied the time dependence of \(W_2\), \(W_3\),
	and \(W_4\) at representative fixed values of \(q\).  The acceptance-dependent
	cumulants, however, are not determined by a single fixed-\(q\) mode.  They are
	obtained by integrating the freezeout Wigner functions over \(q\) with the
	acceptance kernel in Eq.~\eqref{eq:C_N-W_N}.  It is therefore useful to inspect the
	\(q\)-dependence of \(W_N(q,t_f)\) at freezeout, especially inside the physical
	momentum window \([q_{\min},q_{\max}]\) used in the cumulant calculation.
	
\begin{figure}[t]
	\centering
	\includegraphics[width=.32\textwidth]{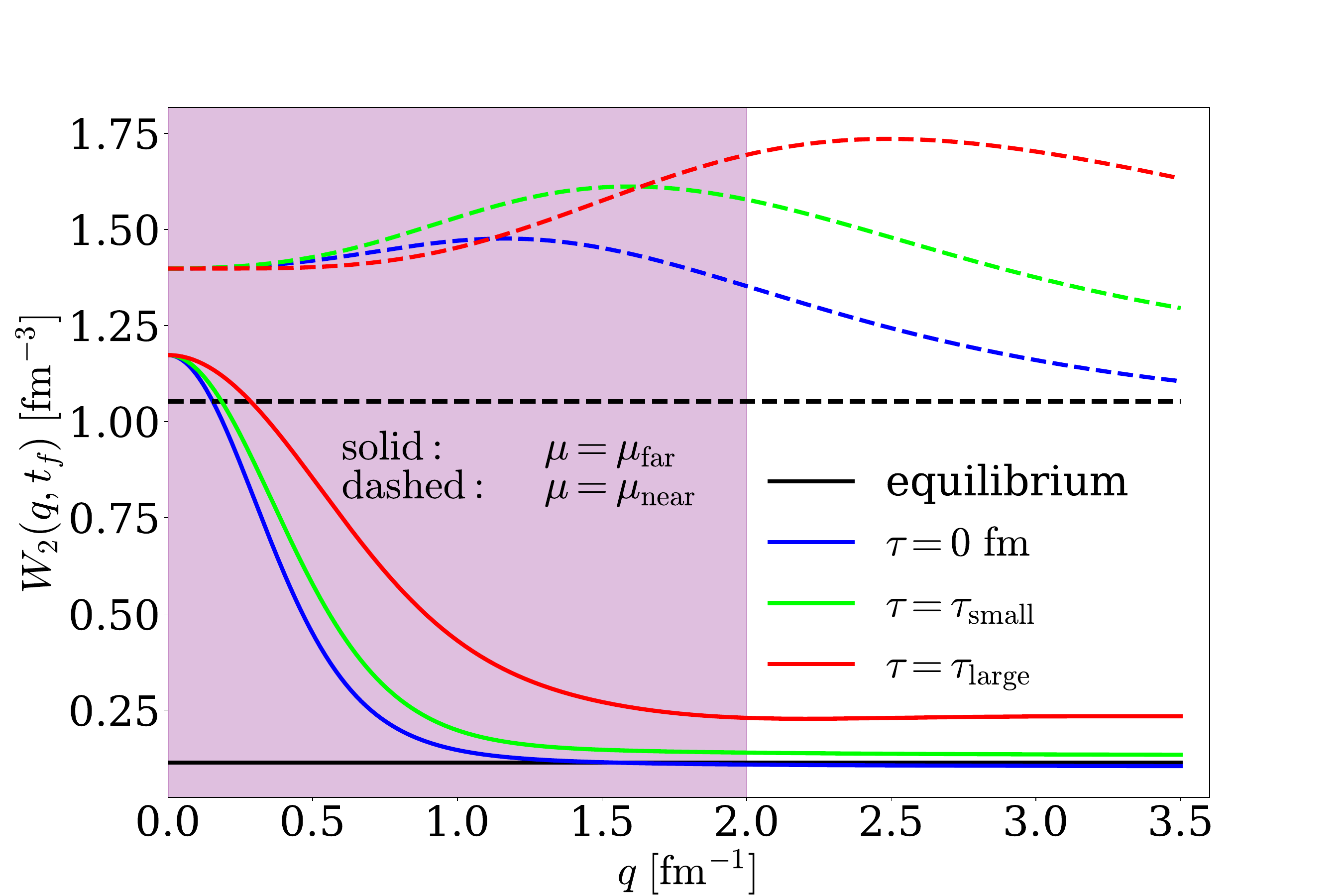}
	\includegraphics[width=.32\textwidth]{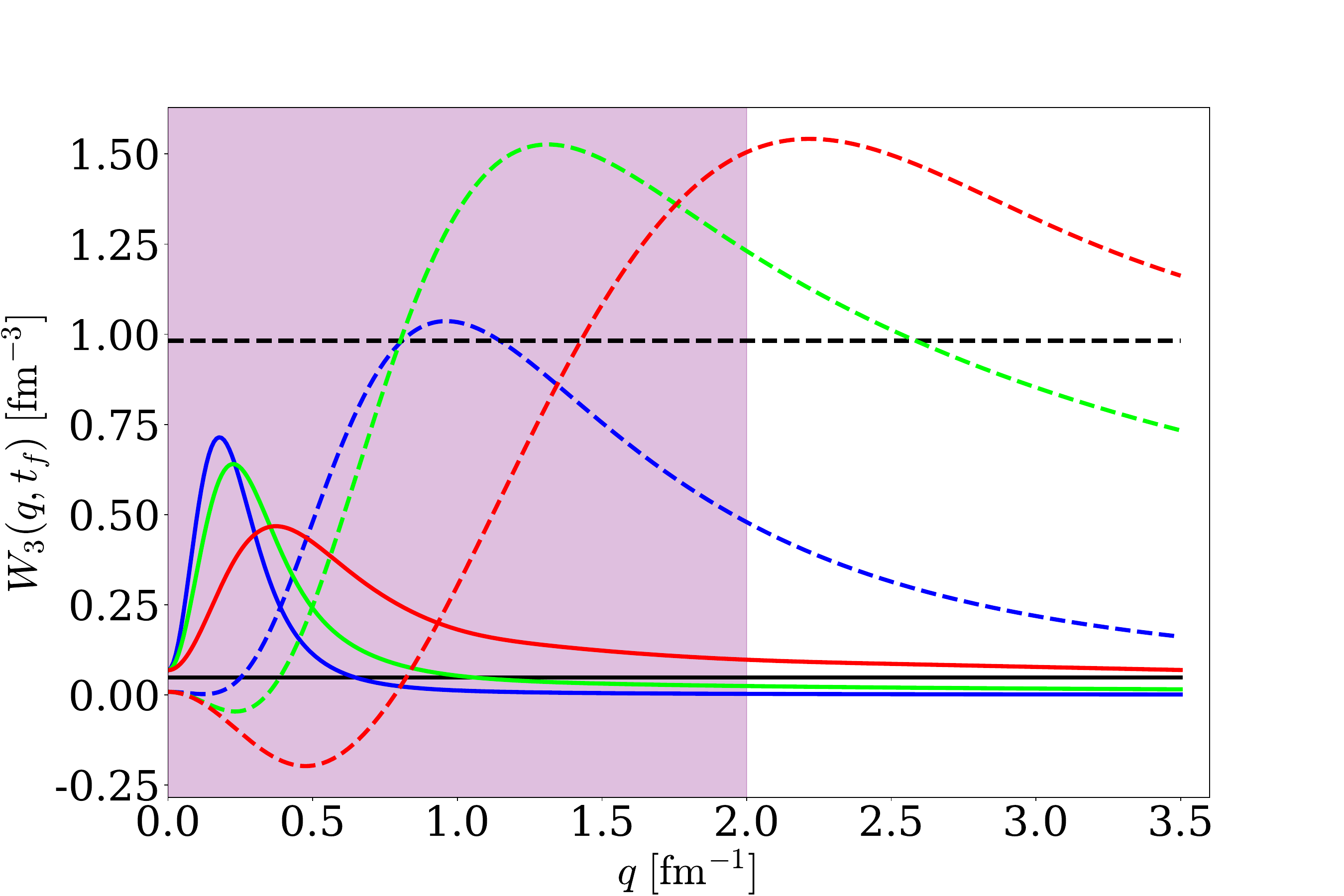}
	\includegraphics[width=.32\textwidth]{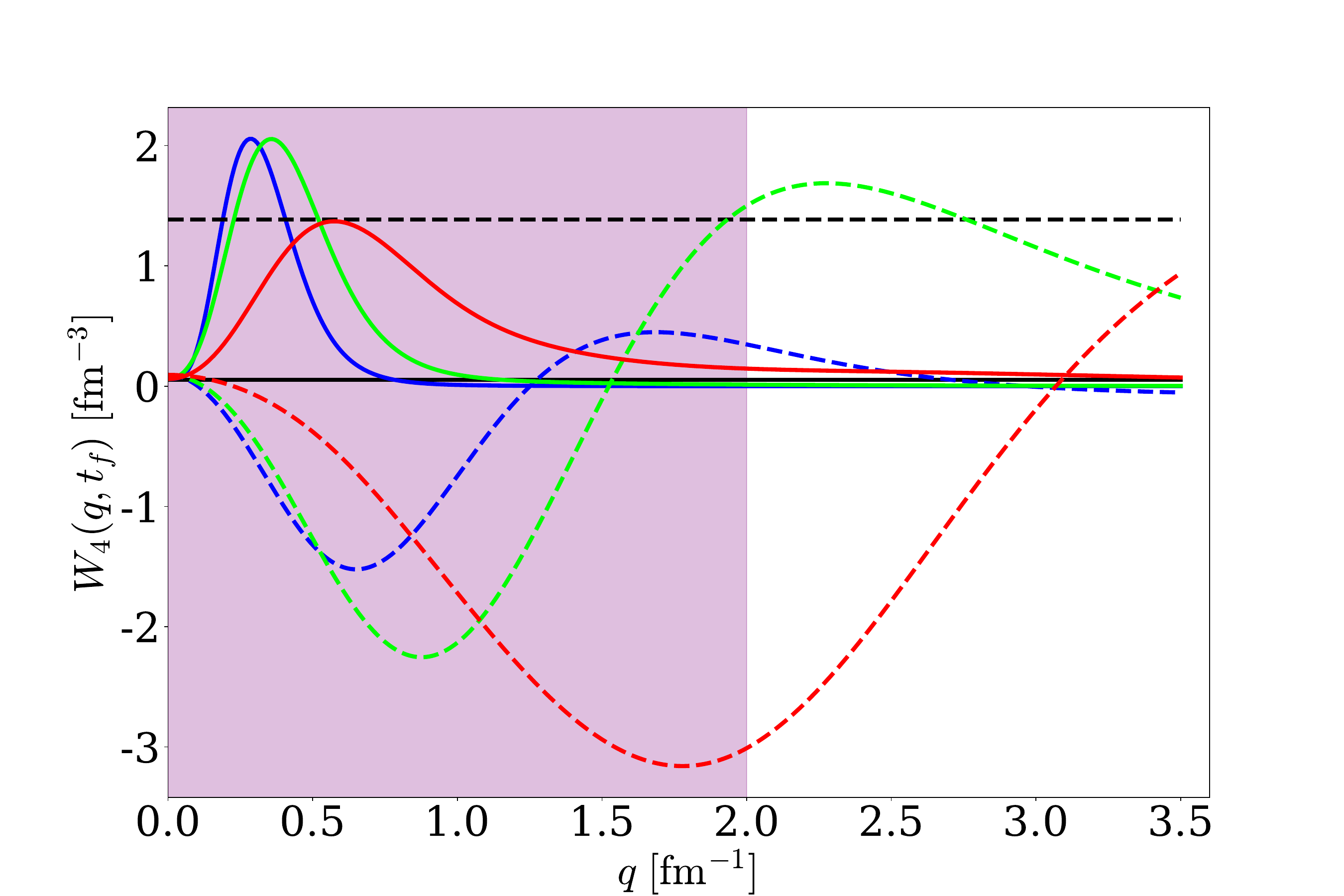}
	\caption{The Wigner functions \(W_{k=2,3,4}(q,t_f)\) as functions of \(q\)
		at freezeout time \(t_f\). The solid and dashed curves correspond to
		evolution trajectories at fixed \(\mu_{\rm far}=0.100~{\rm GeV}\) and
		\(\mu_{\rm near}=0.366~{\rm GeV}\), respectively. The black curves represent
		the corresponding equilibrium values \(W_k^{\rm eq}\). The colored curves
		show \(W_k(q,t_f)\) for different relaxation times, including the Fickian
		limit \(\tau=0\) shown in blue. The magenta region denotes the physical
		momentum window \([q_{\min},q_{\max}]=[0,2]\,{\rm fm}^{-1}\) adopted in
		this work.}
	\label{fig:W234_q}
\end{figure}

	In this subsection we use \(W_2(q,t_f)\) as the simplest example to understand
	the limiting small-\(q\) and large-\(q\) behavior analytically, and then compare
	with the numerical freezeout spectra shown in Fig.~\ref{fig:W234_q}.  The same figure also displays \(W_3(q,t_f)\) and \(W_4(q,t_f)\), which show the stronger sensitivity
	of higher-point correlators to finite current relaxation.

Eliminating \(X_2\) and \(Y_2\) in
Eqs.~\eqref{eq:evo_W2}--\eqref{eq:evo_Y2}, one obtains
\begin{equation}\label{eq:PDE_W2(q)}
	\left[
	\frac{\tau^2}{2}\partial_t^3
	+\frac{3\tau}{2}\partial_t^2
	+\left(1+2\tau\gamma q^2\right)\partial_t
	+2\gamma q^2
	\right]W_2(q,t)=2\lambda q^2\, .
\end{equation}
The two limiting cases of Eq.~\eqref{eq:PDE_W2(q)} clarify the behavior of the
freezeout spectrum shown in Fig.~\ref{fig:W234_q}. In the long-wavelength limit
\(q\to0\), Eq.~\eqref{eq:PDE_W2(q)} reduces to
\begin{equation}\label{eq:PDE_W2_qzero}
	\left(
	\frac{\tau^2}{2}\partial_t^3
	+\frac{3\tau}{2}\partial_t^2
	+\partial_t
	\right)W_2(0,t)=0\, ,
\end{equation}
whose solution is
\begin{equation}\label{eq:W2_qzero_sol}
	W_2(0,t)=W_2^{\rm eq}(0,t_0)
	+\tau\left(
	-C_1-C_2
	+C_1e^{-(t-t_0)/\tau}
	+C_2e^{-2(t-t_0)/\tau}
	\right)\, ,
\end{equation}
where \(C_{1,2}(\tau,t_0)\) are integration constants. Thus the exactly
zero-momentum mode is nondissipative. In the Fickian limit, the freezeout
Wigner function retains its initial value,
\(W_2(0,t_f)\to W_2(0,t_0)\), because the diffusive relaxation rate vanishes as
\(q\to0\). At finite \(\tau\), the auxiliary sector only produces transient
modes with decay times of order \(\tau\); after these transients decay,
\(W_2(0,t)\) again approaches a time-independent constant. With the initial
conditions used in the numerical implementation of Fig.~\ref{fig:W234_q}, this
constant is the same for all finite-\(\tau\) curves, namely
\(W_2^{\rm eq}(0,t_0)\).

The opposite limit is also instructive. Taking \(q\to\infty\) in
Eq.~\eqref{eq:PDE_W2(q)}, and keeping the leading terms in \(1/q\), gives the
asymptotic relaxation-time form
\begin{equation}\label{eq:W2_qinfty}
	\partial_t W_2(\infty,t)
	=-\frac{W_2(\infty,t)-W_2^{\rm eq}(t)}{\tau}\, ,
\end{equation}
where \(W_2^{\rm eq}=\lambda/\gamma=1/\alpha'\). Its solution is
\begin{equation}\label{eq:W2_largeq_sol}
	W_2(\infty,t)
	=e^{-(t-t_0)/\tau}W_2(\infty,t_0)
	+\int_{t_0}^t ds\,\frac{e^{-(t-s)/\tau}}{\tau}
	W_2^{\rm eq}(s)\, .
\end{equation}
For \(\tau=0\), the exponential kernel becomes a delta function and the
large-\(q\) modes track the instantaneous equilibrium value,
\(W_2(\infty,t)=W_2^{\rm eq}(t)\). This is why the blue Fickian curves in
Fig.~\ref{fig:W234_q} approach the black equilibrium lines at large \(q\). At
finite \(\tau\), however, the large-\(q\) modes do not relax instantaneously.
Instead, they follow a retarded average of the equilibrium target over a time
scale \(\tau\). For a slowly varying target, Eq.~\eqref{eq:W2_qinfty} gives
\begin{equation}\label{eq:W2_largeq_gradient}
	W_2(\infty,t)
	\simeq W_2^{\rm eq}(t)-\tau\,\partial_t W_2^{\rm eq}(t)\, .
\end{equation}
Since \(W_2^{\rm eq}=T\chi_2\) decreases near freezeout in our trajectories
(cf. Fig.~\ref{fig:coeff}), one has \(\partial_t W_2^{\rm eq}<0\), and the
finite-\(\tau\) large-\(q\) value lies above the instantaneous-equilibrium
baseline.

For the cumulants, however, the relevant modes are not the asymptotic
\(q\to0\) or \(q\to\infty\) limits, but the modes inside the physical momentum
window shown by the shaded band in Fig.~\ref{fig:W234_q}. The acceptance kernel
gives the largest weight to the low-\(q\) part of this window. Focusing on
\(W_2\), the far-from-critical trajectory shows a relatively smooth dependence
on \(\tau\), because the equilibrium target varies smoothly. Along the
near-critical trajectory, by contrast, the equilibrium target has a pronounced
non-monotonic time dependence and the relaxation of the correlator is less
efficient. As a result, the freezeout spectrum \(W_2(q,t_f)\) develops a
nontrivial dependence on both \(q\) and \(\tau\), which is important for the
behavior of \(C_2\) discussed in Fig.~\ref{fig:Cm_scan}.

The higher-point spectra in Fig.~\ref{fig:W234_q} show the same mechanism in a
more amplified form. Because \(W_3\) and \(W_4\) inherit sharper and
sign-changing equilibrium structures, finite current relaxation produces larger
distortions of their freezeout \(q\)-profiles than in \(W_2\). This anticipates
the stronger finite-\(\tau\) effects in the acceptance-dependent \(C_3\),
\(C_4\), and ratio observables discussed in the next subsection.

It is important to emphasize that our correlator evolution equations apply to $q\lesssim 1/\xi$, consistent with the hydrodynamic regime $\epsilon\sim (q\xi)^3\lesssim 1$. Extension to the scaling regime $q\xi\gtrsim1$ requires resummation and modification as in Model H~\cite{An:2020jjk}. Such a treatment suppresses the large-$q$ critical enhancement relative to the local long-wavelength approximation used here. The acceptance kernel $A_\Delta$ in Eq.~\eqref{eq:C_N-W_N}, on the other hand, predominantly weights modes with $q\lesssim 2\pi/\Delta\sim 2~{\rm fm}^{-1}$, rather than imposing a sharp cutoff. 

Far from the critical point, where $\xi\sim 0.5~{\rm fm}$, most of this momentum range remains within the regime $q \xi\ll 1$. Close to  the critical point, for example for $\xi\sim 3~{\rm fm}$, the controlled long-wavelength regime is restricted to $q\lesssim 0.3~{\rm fm}^{-1}$, while the contribution from larger $q$ requires the aforementioned extension. Since resolving the finite correlation length is expected to reduce the critical contribution from these larger-$q$ modes, the present local treatment may overestimate their contribution to the magnitude of the nonequilibrium cumulants. We nevertheless expect the qualitative finite-memory effects discussed below, which originate from the delayed evolution of the long-wavelength modes, to remain unchanged.

	\subsection{Cumulants in a finite acceptance window}
	\label{subsec:cumulants}
	We now convert the evolved equal-time correlators at freezeout into the
	acceptance-dependent cumulants reported in experiments. In practice, for a rapidity interval of width $\Delta y$, the reduced cumulants $C_N(\Delta\simeq \uptau\Delta y)$ are obtained by evaluating the Wigner functions with the acceptance-kernel formula~\eqref{eq:C_N-W_N} at freezeout time for each background trajectory.
	
	Motivated by experiments such as the BES program at RHIC, we present the freezeout
	cumulants as functions of chemical potential $\mu$ which is inversely related to the beam energy. In our simplified setup, varying $\mu$ moves the system between trajectories that remain far from the critical region and trajectories that pass close to it. This provides a controlled way to quantify how finite current relaxation modifies the observable cumulants, and how this modification is amplified near critical point.
	
\begin{figure}[t]
		\centering
		\includegraphics[width=.32\textwidth]{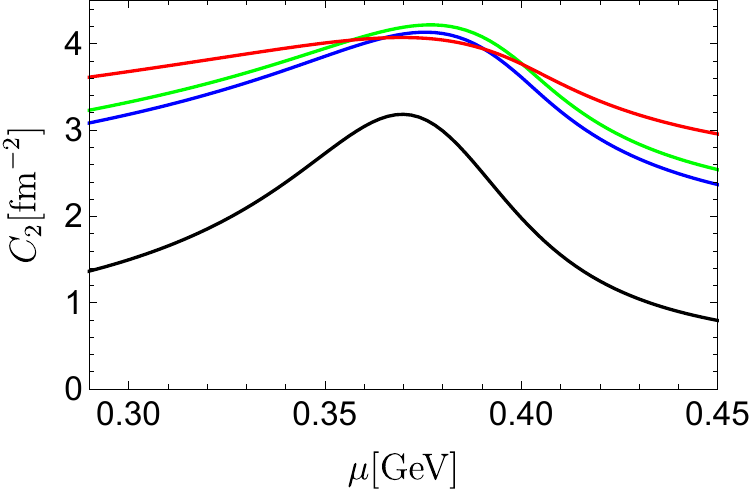}	\includegraphics[width=.32\textwidth]{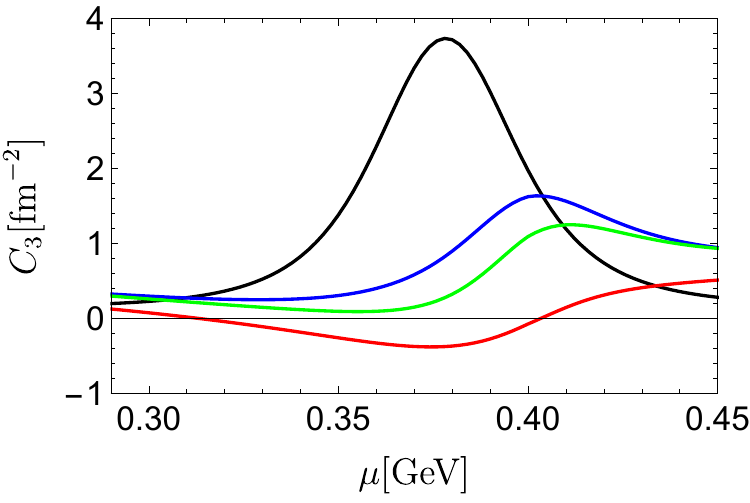}	\includegraphics[width=.32\textwidth]{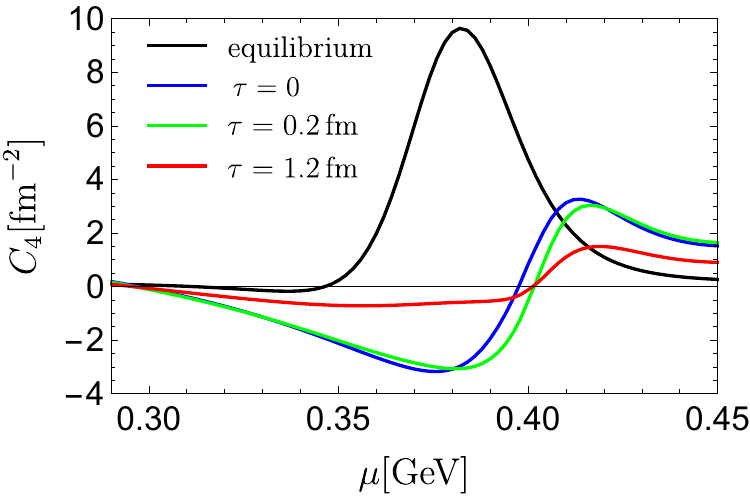}
		\caption{Freezeout acceptance-dependent cumulants $C_2$, $C_3$, and $C_4$ for a representative rapidity window, $\Delta y=0.5$, shown as functions of the trajectory parameter $\mu$ around $\mu_c$. The black curves denote the corresponding instantaneous local-equilibrium
			estimates obtained from the equilibrium correlators at freezeout. The blue curve shows the Fickian diffusion baseline ($\tau=0$), while the green and red curves show MC evolution with
			$\tau=0.2~\mathrm{fm}$ and $\tau=1.2~\mathrm{fm}$, respectively.
			Finite current-relaxation modifies the height, position, and sign structure of the cumulant curves, with the effect becoming more pronounced for higher cumulants.}
		\label{fig:Cm_scan}
	\end{figure}

	Fig.~\ref{fig:Cm_scan} shows the freezeout cumulants $C_2$, $C_3$, and $C_4$ for a representative acceptance width $\Delta y =0.5$, plotted as functions of the trajectory parameter $\mu$. The black curves denote the corresponding instantaneous local-equilibrium estimates, while the colored curves show the cumulants obtained from the dynamical evolution with different MC relaxation times.

Several important features of these plots are as follows. First, the cumulants
should not be read as direct images of a representative fixed-\(q\) curve. They
are acceptance-weighted freezeout integrals over \(W_N(q,t_f)\), and
Eq.~\eqref{eq:C_N-W_N} shows that the acceptance kernel gives the
largest weight to the low-\(q\) part of the physical momentum window. This point
is especially transparent for \(C_2\), whose kernel is positive: the ordering of
the \(C_2\) curves in Fig.~\ref{fig:Cm_scan} follows the ordering of the
corresponding low-\(q\) modes of \(W_2(q,t_f)\) inside the physical momentum
window shown in Fig.~\ref{fig:W234_q}. The detailed ordering of the \(C_2\)
curves is not universal, since it depends on the freezeout condition, on when
the trajectory crosses the region where the susceptibilities vary rapidly, and
on the chosen acceptance window. 
	
	Second, the effect is more pronounced for $C_3$ and $C_4$, because the relevant equilibrium structures are sharper and sign-changing. Finite current relaxation can then shift extrema, change magnitudes, and modify the sign pattern.

	The smoother behavior of the cumulant curves in Fig.~\ref{fig:Cm_scan} should not be interpreted as an absence of the underdamped MC response presented in the fixed-\(q\) curves for correlators shwon in Figs.~\ref{fig:W_2_far}, \ref{fig:W_2_near}, ~\ref{fig:W_34_far} and \ref{fig:W_34_near}. Considering Eq.~\eqref{eq:C_N-W_N},  since the MC oscillation frequency depends on \(q\), different momentum modes accumulate different phases before freezeout and therefore add only partially coherently after acceptance integration. The conversion to \(C_n(\Delta y)\) therefore smooths the fixed-$q$ oscillations, while retaining the net finite-\(\tau\) effect in the height, position, and the sign structure of the cumulant curves. 
	
	\begin{figure}[t]
		\centering
		\includegraphics[width=.45\textwidth]{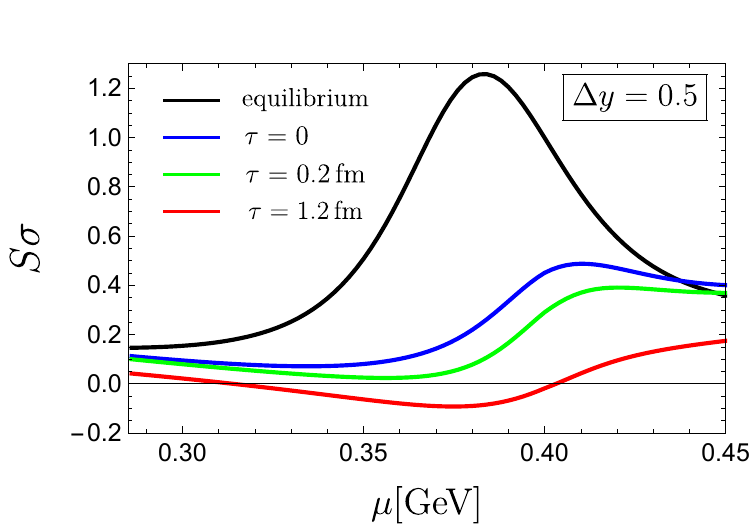}	\,\,\includegraphics[width=.44\textwidth]{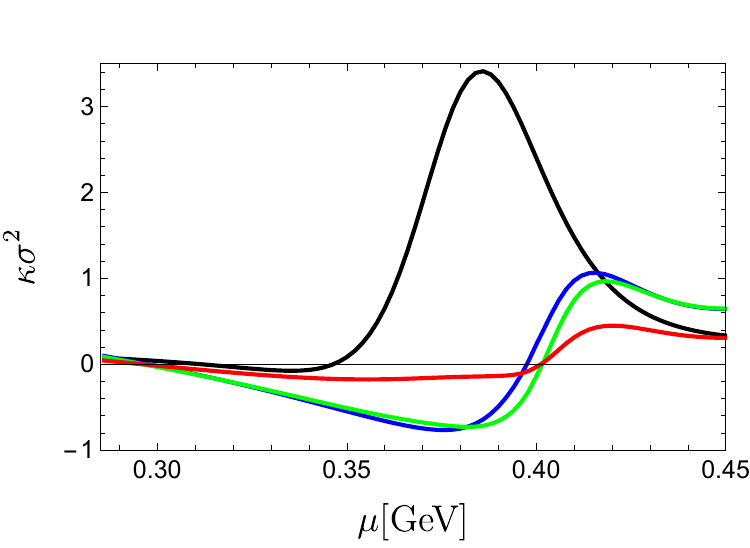}
		\caption{		
			Freezeout ratio observables \(S\sigma \equiv C_3/C_2\) and
			\(\kappa\sigma^2 \equiv C_4/C_2\) at fixed acceptance width
			\(\Delta y=0.5\), shown as functions of \(\mu\). Color conventions are the
			same as in Fig.~\ref{fig:Cm_scan}; black for the instantaneous local-equilibrium estimate, blue for the Fickian diffusion baseline \((\tau=0)\), and green/red for MC
			evolution with \(\tau_{\rm small}=0.2~{\rm fm}\) and
			\(\tau_{\rm large}=1.2~{\rm fm}\). Finite current relaxation suppresses the
			near-critical enhancement of the ratio observables relative to the
			instantaneous-equilibrium estimate; the effect is especially visible in
			\(\kappa\sigma^2\), where it shifts the minimum and can modify the sign
			relative to the Fickian baseline over part of the scanned range.}
		\label{fig:skewness_kurtosis}
	\end{figure}
	
	To display these effects in the commonly used ratio observables, Fig.~\ref{fig:skewness_kurtosis} shows	\(S\sigma \equiv C_3/C_2\) and \(\kappa\sigma^2 \equiv C_4/C_2\) at	\(\Delta y=0.5\).  The figure gives a simple phenomenological message: finite current relaxation suppresses the near-critical enhancement of these ratio observables relative to the instantaneous local-equilibrium estimate.  The suppression is visible in \(S\sigma\), through both the height and the position of its peak, and is more pronounced in \(\kappa\sigma^2\), where finite
	\(\tau\) reduces the peak structure, shifts the minimum, delays the recovery,
	and can modify the sign relative to the \(\tau=0\) Fickian baseline over part
	of the scanned range.  Thus the experimentally relevant skewness and kurtosis
	ratios retain clear finite-memory imprints even after the fixed-\(q\)
	oscillatory response has been smoothed by acceptance integration.\\

	\section{Conclusion and outlook}
	\label{sec:conclusion}
Diffusion is one of the most widely studied transport processes in physics.
In this work, we studied how finite current relaxation modifies the stochastic
diffusion of a conserved charge near the QCD critical point. In Fickian
diffusion, the diffusive current is slaved instantaneously to the local
thermodynamic force, \(J^i=-\lambda\nabla^i\alpha\). We instead used
Maxwell--Cattaneo theory as a minimal finite-memory extension, in which the
diffusive current is promoted to an independent dynamical variable relaxing
toward its Fickian value over a time scale \(\tau\). In this stochastic enlarged
variable space, we derived closed evolution equations for the equal-time
Wigner functions \(W_N\) with \(N=2,3,4\) associated with charge correlators, and followed their
evolution along representative trajectories in the QCD phase diagram. After
freezeout, these correlators were converted into acceptance-dependent cumulants
relevant for heavy-ion fluctuation measurements.

	Our main result is that finite current relaxation introduces a genuine memory effect beyond the Fickian diffusion baseline.  Even in Fickian diffusion the correlators need not track the evolving local-equilibrium target  instantaneously; finite $\tau$ adds an additional response time  controlled by the competition between diffusion and the relaxation time $\tau$. This effect becomes phenomenologically relevant when the characteristic MC scale $q_\star\sim 1/(2\sqrt{\tau \gamma})$ overlaps the physical momentum window entering the acceptance-dependent cumulants. In that regime, the correlators can be shifted relative to the diffusion baseline and, for sufficiently large $\tau$, can show a weakly underdamped response.
	
	Our second main result is that this finite-relaxation effect becomes more visible near the critical region. There, the equilibrium susceptibilities vary rapidly along the trajectory, while the relaxation of the correlators becomes less
	efficient because of critical slowing down. As a result, the
	deviations from the Fickian diffusion baseline become much larger for near-critical
	trajectories than for trajectories that remain far from critical point.

Our third result is that finite current memory does not act merely as an
overall suppression or enhancement of the cumulants.  For \(C_2\), the acceptance kernel is positive and gives its largest weight to the low-\(q\) sector of \(W_2(q,t_f)\). Consequently, \(C_2(\mu)\) is controlled by the finite-\(\tau\) modification of the low-\(q\)
freezeout spectrum, rather than by a single representative fixed-\(q\) curve.
Since this low-\(q\) spectrum depends on the freezeout time and on the chosen
acceptance window, the detailed shape of \(C_2(\mu)\) should not be regarded as
universal. For \(C_3\), \(C_4\) the delayed response is more visible: finite
\(\tau\) can suppress the near-critical enhancement, shift extrema, and modify the sign over part of the scanned range.

The phenomenological implication is that the observed critical signal is not
determined by the equation of state alone. Even with the same equilibrium
susceptibilities, the nonequilibrium evolution of fluctuations can lead to
different predictions for the height, location, and sign structure of the
freezeout cumulants. Already within Fickian diffusion, fluctuation correlators
need not follow their instantaneous equilibrium values, and the near-critical
enhancement of the experimentally relevant ratio observables can be weakened.
Finite current memory provides an additional source of delay and can further
suppress this near-critical enhancement in the ratio observables.
Thus, current relaxation should be viewed as a possible dynamical source of
systematic uncertainty in interpreting cumulant measurements near the QCD
critical point.

	Several limitations of the present analysis should be kept in mind. First, the background evolution is prescribed rather than solved self-consistently; we use representative cooling trajectories and keep $\mu$ fixed along each trajectory in order to isolate finite-relaxation effects in a controlled setting. 
     Second, our effective description is one-dimensional and restricted to the long-wavelength regime where the local hydrodynamic description applies; the full coupling to an explicitly expanding, spatially inhomogeneous background and the extension into the critical scaling regime are not included. Third, MC theory is employed here as a minimal model of memory effects, rather than as a first-principles description of critical dynamics. Fourth, fluctuations of the energy-momentum sector, including entropy and velocity fluctuations as studied in Refs.~\cite{An:2024nvr,An:2026glk,Basar:2026fol}, are not taken into account. Finally, although we convert freezeout correlators into reduced acceptance-dependent cumulants, we do not yet implement a full particlization map to experimentally measured hadron cumulants.
	
	These limitations naturally suggest several directions for future work. A key next step is to embed the present mechanism in a more realistic expanding background that includes energy and momentum fluctuations~\cite{An:2023yfq}, relaxes the constant-$\mu$ trajectory assumption, and allows for a systematic study of the dependence on momentum and acceptance windows. It will also be important to connect the freezeout correlators more directly to experimentally measured hadronic observables through a full particlization procedure. In parallel, the minimal MC description used here should be compared with more microscopic realizations of non-diffusive critical dynamics. By isolating, within a controlled framework, how finite current relaxation can leave observable imprints on fluctuation cumulants near the critical region, the present results provide a useful step toward these broader goals.

	As broader directions, it would be interesting to explore whether similar finite-memory effects appear in the Brownian motion of heavy quarks in the QGP near the critical point~\cite{Oei:2024pva,Rajagopal:2025rxr,Liu:2021dpm}, as well as the fluctuation dynamics near a O(4) chiral critical point~\cite{Schlichting:2019tbr,Florio:2025zqv}.  It may also be useful to investigate whether analogous ideas apply to diffusive processes in neural-network dynamics and artificial intelligence~\cite{Aarts:2025gyp, Zhu:2026dmc,raissi2019physics}, where memory effects and non-Fickian transport can play an important role.

	\section*{Acknowledgment}
	We thank Matteo Baggioli, Wei-jie Fu, Michal Heller, Xiaofeng Luo, Huichao Song, Mikhail Stephanov and Yi Yin for helpful discussions. N.A. acknowledges support from the National Natural Science Foundation of China under Grant No.~12575142, and from the “Double First-Class” start-up funding of Lanzhou University, China, under Grant No.~561119208. X.A. is supported by the European Research Council (ERC) under the European Union’s EU Horizon 2020 research and innovation program (Grant No.~101089093/project acronym: High-TheQ). S.W. is supported by the National Natural Science Foundation of China under grant No.~12305143 and the Fundamental Research Funds for the Central Universities at Dalian University of Technology.
	
	\appendix
	
	\section{Integrating out Hydro\texorpdfstring{$+$}{+} mode and the emergence of a dynamical current}
	\label{app:hydroplus_mc}
	In this Appendix we briefly justify the statement made in the Introduction that integrating out an additional slow mode produces memory in the constitutive relation. We use the Hydro\(+\) setup as a concrete illustration~\cite{Stephanov:2017ghc,Du:2021zqz}.
	
	In Hydro\(+\), one enlarges the local thermodynamic description by including a slow variable \(\phi\), with generalized entropy		
	\begin{equation}\label{eq:ds(+)}
		ds_{(+)}=\beta_{(+)}\,d\varepsilon-\alpha_{(+)}\,dn-\pi\,d\phi\,.
	\end{equation}		
	The quantity \(\pi(\varepsilon,n,\phi)\) is the thermodynamic force conjugate to \(\phi\). In the equilibrium,
	\begin{equation}\label{eq:phi^eq}
		\phi=\phi^{\rm eq}(\varepsilon,n)\,,
		\qquad
		\pi\bigl(\varepsilon,n,\phi^{\rm eq}(\varepsilon,n)\bigr)=0\,.
	\end{equation}		
	Thus \(\pi\) does not relax toward a nonzero equilibrium value; rather, it vanishes in equilibrium and becomes nonzero only when the slow mode lags behind its instantaneous equilibrium value.
	
	To see how the slow-mode is driven, let us see first how the equilibrium value \(\phi^{\rm eq}(\varepsilon,n)\) changes along the flow. Using the ideal conservation laws,		
	\begin{equation}\label{eq:De-Dn}
		D\varepsilon=-(\varepsilon+p)\,\Theta\,,
		\qquad
		Dn=-n\,\Theta\,,
		\qquad{\rm where}\qquad
		D\equiv u\cdot\partial\,, \qquad \Theta\equiv\partial\cdot u\,, 
	\end{equation}		
	one finds		
	\begin{equation}\label{eq:Dphi^eq}
		D\phi^{\rm eq}
		=
		\phi^{\rm eq}_\varepsilon D\varepsilon
		+
		\phi^{\rm eq}_nDn
		\nonumber =-[(\epsilon+p)\phi^{\rm eq}_\varepsilon+n\phi^{\rm eq}_n]\Theta\equiv
		-\mathcal A_\Theta\,\Theta\,.
	\end{equation}
	Therefore, even if initially \(\phi=\phi^{\rm eq}\), a nonzero expansion rate \(\Theta\) shifts the instantaneous equilibrium. If \(\phi\) is slow, it cannot follow this shift immediately, and a nonzero force \(\pi\) is generated.
	
	Defining the deviation from the equilibrium as
	$\varphi \equiv \phi-\phi^{\rm eq}(\varepsilon,n),$
	the linear relaxation of the slow mode can be written as		
	\begin{equation}\label{eq:Dphi}
		D\phi=-\Gamma_\phi\,\varphi\,.
	\end{equation}		
	Using \eqref{eq:De-Dn}, and \eqref{eq:Dphi}, we obtain		
	\begin{equation}\label{D_var_phi}
		D\varphi
		=
		D\phi-D\phi^{\rm eq}
		=
		-\Gamma_\phi\,\varphi
		+\mathcal A_\Theta\,\Theta\, .
	\end{equation}		
	On the other hand, near equilibrium, \(\pi\) is linear in \(\varphi\),
	\begin{equation}\label{eq:pi=Kphi}
		\pi=\mathcal K\,\varphi\,,
	\end{equation}		
	so Eq.~\eqref{D_var_phi} becomes
	\begin{equation}\label{eq:Dpi}
		\tau_\pi\,D\pi+\pi=\kappa_\Theta\,\Theta\,,
		\qquad
		\tau_\pi\equiv \Gamma_\phi^{-1}\,,
		\qquad
		\kappa_\Theta\equiv \tau_\pi \mathcal K \mathcal A_\Theta\,.
	\end{equation}		
	This is the local relaxation equation for the Hydro\(+\) force \(\pi\).
	
	Passing to frequency space, we have		
	\begin{equation}	\label{eq:pi(omega)}
		\pi(\omega)=\frac{\kappa_\Theta}{1-i\omega\tau_\pi}\,\Theta(\omega) \,.
	\end{equation}		
	Equivalently, in the time domain, $\pi$ is a non-local function of $\Theta$		
	\begin{equation}\label{eq:pi(t)}
		\pi(t)
		=
		\kappa_\Theta
		\int_{-\infty}^{t}dt'\,
		\frac{e^{-(t-t')/\tau_\pi}}{\tau_\pi}\,
		\Theta(t')\, .
	\end{equation}		
	Thus, once the slow mode is integrated out, \(\pi\) depends on the past history of the fluid expansion.
	
	The Hydro\(+\) constitutive relation for the diffusive current contains the spatial gradient of the force $\pi$		
	\begin{equation}	\label{eq:Jhydroplus}
		\Delta J^\mu
		=
		-\lambda_{\alpha\alpha}\,\nabla^\mu \alpha
		-\lambda_{\alpha\pi}\,\nabla^\mu \pi\,.
	\end{equation}		
	Substituting Eq.~\eqref{eq:pi(omega)}, we obtain		
	\begin{equation} 	\label{eq:Jomega}
		\Delta J^\mu(\omega)
		=
		-\lambda_{\alpha\alpha}\,\nabla^\mu \alpha(\omega)
		-\lambda_{\alpha\pi}\,
		\frac{\kappa_\Theta}{1-i\omega\tau_\pi}\,
		\nabla^\mu \Theta(\omega) \,.
	\end{equation}		
	In the time domain this becomes		
	\begin{equation}	\label{eq:Jmemory}
		\Delta J^\mu(t)
		=
		-\lambda_{\alpha\alpha}\,\nabla^\mu \alpha(t)
		-\lambda_{\alpha\pi}\kappa_\Theta
		\int_{-\infty}^{t}dt'\,
		\frac{e^{-(t-t')/\tau_\pi}}{\tau_\pi}\,
		\nabla^\mu \Theta(t') \,.
	\end{equation}	
	This equation illustrates how the current depends on the past history of the fluid expansion.
	Eq.~\eqref{eq:Jmemory} may be rewritten in an equivalent local form by introducing an auxiliary, memory-carrying contribution to the dissipative current,	
	\begin{equation}\label{eq:DeltaJ}
		\Delta J^\mu
		=
		-\lambda_{\alpha\alpha}\nabla^\mu \alpha
		+
		j_{\rm mem}^\mu\, .
	\end{equation}	
	Comparing with Eq.~\eqref{eq:Jmemory}, one identifies	
	\begin{equation}	\label{eq:J_mem}
		j_{\rm mem}^\mu(t)
		=
		-\lambda_{\alpha\pi}\kappa_\Theta
		\int_{-\infty}^{t}dt'\,
		\frac{e^{-(t-t')/\tau_\pi}}{\tau_\pi}\,
		\nabla^\mu\Theta(t') \,.
	\end{equation}	
	This is exactly equivalent to the local relaxation equation	
	\begin{equation}\label{eq:DJ_mem}
		\tau_\pi D j_{\rm mem}^\mu + j_{\rm mem}^\mu
		=
		-\lambda_{\alpha\pi}\kappa_\Theta\,\nabla^\mu\Theta \,.
	\end{equation}	
	Thus, after integrating out the slow Hydro$+$ mode, the constitutive relation may be written either in nonlocal form, Eq.~\eqref{eq:Jmemory}, or equivalently in local form by promoting the memory-carrying part of the current to an additional dynamical variable obeying a relaxation-type equation.
	
	In the Hydro$+$ example considered here, the slow mode is sourced by the scalar expansion rate $\Theta$, so the delayed contribution naturally appears in the scalar channel. The MC theory used in the main text should be viewed as the minimal analogous construction in the diffusion sector, in which the delayed contribution is assigned directly to the diffusive current itself:	
	\begin{equation}\label{eq:J_MC_app}
		\tau D\Delta J^\mu + \Delta J^\mu
		=
		-\gamma\,\nabla^\mu \delta n \,.
	\end{equation}	
	Equation~\eqref{eq:J_MC_app} is therefore not the literal Hydro$+$ constitutive relation above, but the minimal local realization of the same general idea: a delayed response encoded as a dynamical current.

	\section{Derivation of the KMS condition in MC theory}
	\label{app:KMS}
	In this appendix, we derive the constraints on the transport coefficients in the
	MC stochastic equations by requiring that the dynamics admits the
	expected Gibbs equilibrium distribution $P_{\rm eq}$~\cite{An:2020vri,Gavassino:2024vyu}:
	\begin{equation}\label{eq:S_n}
		\log P_{\rm eq}[n]=S[n]=\int_{\mathbf x} (s(n)+\bar\alpha n)\approx S[\langle n\rangle]-\frac{1}{2}\int_{\mathbf x}\chi_n^{-1}\delta n^2+\dots\,,
	\end{equation}
	where $S[n]$ is the entropy functional of $n$, $\bar\alpha$ is the Lagrange multiplier for the total charge conservation, and $\chi_n^{-1}=\alpha'=-S''$ is the charge susceptibility. Near equilibrium where entropy maximizes, the linear term in the $\delta n=n-\langle n\rangle$ expansion vanishes since $\delta S[n]/\delta n=S'=-(\alpha-\bar\alpha)\simeq0$, while the quadratic term implies an Ornstein–Uhlenbeck process for charge $n$ with stationary variance 
	\begin{equation}\label{eq:dndn}
		\langle \delta n(\mathbf x_1)\delta n(\mathbf x_2)\rangle_{\rm eq}= \chi_n\delta^{(3)}(\mathbf x_1-\mathbf x_2) \quad{\rm where}\quad \chi_n=1/\alpha'\,.
	\end{equation}
	
	We now require the stationary distribution for MC theory to have the extended Gibbs form
	\begin{equation}\label{eq:P-Bayes}
		P_{\rm eq}[n,J]=\,P_{\rm eq}[n]\,P_{\rm eq}[J|n]\propto e^{S[n,J]}\,,
	\end{equation}
	such that for fixed $n$ (such that $\nabla\alpha=0$), Eqs.~\eqref{eq:MC} reduce to an
	Ornstein--Uhlenbeck (Gaussian) process for $J_i$ with stationary variance
	\begin{equation}\label{eq:dJdJ}
		\langle \delta J_i(\mathbf x_1)\delta J_j(\mathbf x_2)\rangle_{\rm eq}=\chi_g\delta_{ij}\delta^{(3)}(\mathbf x_1-\mathbf x_2) \quad{\rm where}\quad \chi_g=\frac{\sigma T}{\tau}\,.
	\end{equation}
	We keep the density entropy functional $S[n]$ in its thermodynamic form, without explicitly expanding it in powers of $\delta n$ but take the current sector to be Gaussian at fixed $n$.
	Thus the entropy functional depending on both density $n=\langle n\rangle+\delta n$ and current $J$ reads
	\begin{equation}\label{S_nj}
		\log P_{\rm eq}[n,J]=S[n,J]=S[n]-\frac{1}{2}\int_{\mathbf x}\chi_g^{-1}(\delta J)^2\,.
	\end{equation}
	Here $\langle J\rangle=0$ in equilibrium, so $J=\delta J$, and $\chi_g$ is the current susceptibility that can be understood as the mass terms for the spatial components of the U(1) gauge field~\cite{An:2025ils,Ahn:2025odk}. In Eq.~\eqref{S_nj} together with Eq.~\eqref{eq:S_n}, both $\chi_n$ and $\chi_g$ may depend on $n$, thus the distribution is Gaussian in $J$ for fixed $n$, but it is not purely quadratic in the joint variables $(n,J)$ if $\chi_g$ depends on $n$.  In particular, expanding $\chi_g^{-1}(n)$ generates the mixed cubic term $-\frac12(\chi_g^{-1})\delta n\,(\delta J)^2$.
	
	From Eq.~\eqref{S_nj} one immediately obtains
	\begin{subequations}
		\begin{align}\label{eq:dSdJn}
			\frac{\delta S[n,J]}{\delta J_i}
			&=-\chi_g^{-1}\,\delta J_i\,,\\
			\frac{\delta S[n,J]}{\delta n}
			&=-(\alpha-\bar\alpha) - \frac12 (\chi_g^{-1})'(\delta J)^2\,.
		\end{align} 
	\end{subequations}
	The corresponding functional Fokker--Planck equation for $P[n,J]$ reads~\cite{An:2020vri}
	\begin{equation}  	\label{FP}
		\partial_t P=
		\int_{\mathbf x}\left[-\frac{\delta}{\delta n}(F_nP)
		-\frac{\delta}{\delta J_i}(F_iP)
		+\frac{\delta^2}{\delta n\delta n}(DP)+\frac{\delta^2}{\delta J_i\delta J_j}(D_{ij}P)\right]\,,
	\end{equation}
	where the drift force terms ($F,F_i$) and noise terms ($D,D_{ij}$) read
	\begin{equation} 	\label{eq:F-D}
		F=-\nabla\cdot J\,,
		\qquad
		F_i=-\frac{1}{\tau}(J_i+\lambda\nabla_i\alpha)\,,
		\qquad D=0\,,\qquad
		D_{ij}=\frac{\chi_g}{\tau}\delta_{ij}\,.
	\end{equation}
	Substituting $P_{\rm eq}\propto e^S$ into Eq.~\eqref{FP}, the Ornstein--Uhlenbeck drift
	$-J_i/\tau$ and the noise term cancel identically by construction of
	Eq.~\eqref{S_nj}. After a short calculation and integration by parts, the only
	remaining contributions are
	\begin{equation}\label{eq:FP_eq}
		\partial_t P_{\rm eq}
		=-\int_{\mathbf x}\,\left[
		\left(1-\frac{\lambda}{\tau\chi_g}\right)
		\alpha
		+\frac12\,(\chi_g^{-1})'\,J^2\right](\nabla\cdot J)\,P_{\rm eq}=0\,.
	\end{equation}
	Since $\alpha$ and $J^2$ are independent for generic field configurations,
	stationarity requires both coefficients to vanish:
	\begin{equation}\label{eq:conds-FP}
		1-\frac{\lambda}{\tau\chi_g}=0,
		\qquad
		(\chi_g^{-1})'=0\,.
	\end{equation}
	Using $\chi_g=\sigma T/\tau$, we obtain the Wiedemann–Franz law in Eq.~\eqref{eq:KMS}, which is precisely the
	fluctuation--dissipation/KMS constraints quoted in the main text. As an immediate
	corollary, if $\tau$ is taken constant, then $\lambda$ must also be constant, and
	so is $\sigma$. See Ref.~\cite{Jain:2023obu} for a different derivation of the constraints given by Eqs.~\eqref{eq:conds-FP}.
	
	\section{Fickian limit of the MC Fokker--Planck equation}
	Let us briefly show how the usual Fick stochastic theory is recovered from the
	MC Fokker--Planck equation in the limit $\tau\to 0$. This limit should not be
	taken at fixed $J_i$, since $J_i$ becomes a fast variable. Instead, one should
	first integrate out the current and consider the marginal probability
	distribution
	\begin{equation}
		P[n]\equiv \int {\cal D}J\, P[n,J] .
	\end{equation}
	We also define the first and second current moments
	\begin{equation}
		{\cal J}_i(x)\equiv \int {\cal D}J\, J_i(x)P[n,J],
		\qquad
		{\cal Q}_{ij}(x,y)\equiv \int {\cal D}J\,J_i(x)J_j(y)P[n,J].
	\end{equation}
	Integrating Eq.~\eqref{FP} over $J_i$, the functional derivatives with respect to
	$J_i$ vanish by parts, and one obtains
	\begin{equation}
		\partial_t P[n]
		=
		\int_{\mathbf x}
		\frac{\delta}{\delta n(x)}
		\nabla_i {\cal J}_i(x).
		\label{eq:PF_marginal}
	\end{equation}

	The equation for ${\cal J}_i$ is obtained by multiplying Eq.~\eqref{FP} by
	$J_i(x)$ and integrating over $J_i$. This gives
	\begin{equation}	\label{Jmoment}
		\tau\partial_t {\cal J}_i(x)
		=
		-{\cal J}_i(x)-\lambda\nabla_i\alpha(x)P
		+
		\tau
		\int_{\mathbf y}
		\frac{\delta}{\delta n(y)}
		\nabla_{y,j}{\cal Q}_{ij}(x,y).
	\end{equation}
	
	In the fast-current limit, the conditional distribution of $J_i$ at fixed $n$
	rapidly approaches the local Gaussian measure
	\begin{equation}
		P[J|n]\propto
		\exp\left[
		-\frac{1}{2\chi_g}\int_{\mathbf x}
		\left(J_i-\bar J_i\right)^2
		\right],
		\qquad
		\bar J_i=-\lambda\nabla_i\alpha ,
		\qquad
		\chi_g=\frac{\lambda}{\tau}.
	\end{equation}
	Therefore
	\begin{equation}
		\int {\cal D}J\,J_i(x)J_j(y)P[J|n]
		=
		\bar J_i(x)\bar J_j(y)
		+
		\chi_g\delta_{ij}\delta^{(3)}(x-y).
	\end{equation}
	Since $\bar J_i$ is finite as $\tau\to0$, while
	$\chi_g=\lambda/\tau$, the leading contribution to the second moment is
	\begin{equation}
		{\cal Q}_{ij}(x,y)
		=
		\frac{\lambda}{\tau}\delta_{ij}\delta^{(3)}(x-y)P+\mathcal O(1),
	\end{equation}
	where we have used $P[n,J]=P[n]P[J|n]$.
	Thus the last term in Eq.~\eqref{Jmoment} remains finite:
	\begin{equation}
		\tau{\cal Q}_{ij}(x,y)
		\longrightarrow
		\lambda\,\delta_{ij}\delta^{(3)}(x-y)P .
	\end{equation}
	Neglecting the $\tau\partial_t{\cal J}_i$ term in the fast-current limit, one
	finds, for constant $\lambda$,
	\begin{equation}	\label{Jmoment_Fick}
		{\cal J}_i(x)
		=
		-\lambda\nabla_i\alpha(x)P
		-\lambda\nabla_i\frac{\delta P}{\delta n(x)}
		+\mathcal O(\tau).
	\end{equation}
	Substituting Eq.~\eqref{Jmoment_Fick} into Eq.~\eqref{eq:PF_marginal}, we
	obtain the Fick Fokker--Planck equation
	\begin{equation}	\label{eq:Fick_FP}
		\partial_t P[n,t]
		=
		-\int_{\mathbf x}
		\frac{\delta}{\delta n(x)}
		\nabla_i
		\left[
		\lambda\nabla_i\alpha(x)P
		+
		\lambda\nabla_i\frac{\delta P}{\delta n(x)}
		\right].
	\end{equation}
	This is precisely the Fokker--Planck equation associated with the stochastic
	Fick constitutive relation
	\begin{equation}
		J_i=-\lambda\nabla_i\alpha+\zeta_i,
		\qquad
		\langle \zeta_i(x,t)\zeta_j(y,t')\rangle
		=
		2\lambda\,\delta_{ij}\delta^{(3)}(x-y)\delta(t-t').
	\end{equation}
	Therefore the apparent singularity of the MC Fokker--Planck equation is not a
	pathology. In the limit $\tau\to0$, the equal-time current fluctuation
	$\chi_g=\lambda/\tau$ diverges, but its correlation time is of order $\tau$.
	Their product remains finite and gives the white stochastic flux of Fick
	diffusion.

	\section{Equilibrium correlators}
	\label{app:equilibrium}
	
	In this Appendix we rewrite the equilibrium measure directly in terms of MC variables $n$ and $g$, specializing to the simple case used in the paper, $\tau = \mathrm{const}$ and $\lambda = \mathrm{const}$. By the KMS condition derived in Appendix~\ref{app:KMS}, this also implies $\sigma=\mathrm{const}$. For constant coefficients, Appendix~\ref{app:KMS} gives the equilibrium entropy functional \eqref{S_nj}.
	
	In Fourier space we decompose the current into longitudinal and transverse parts,
	\begin{equation}\label{eq:J_decompose}
		J_i(\mathbf q)=J_i^{L}(\mathbf q)+J_i^{T}(\mathbf q)\,,
		\qquad
		q_i J_i^{T}(\mathbf q)=0\,.
	\end{equation}
	The continuity equation implies
	\begin{equation}\label{eq:g-J^L}
		g_{\mathbf q}=\partial_t n_{\mathbf q}=-i q_i J_i(\mathbf q)=-i q_i J_i^{L}(\mathbf q)\,,
	\end{equation}
	implying that
	\begin{equation}\label{eq:J^L(q)}
		J_i^{L}(\mathbf q)= i\,\frac{q_i}{q^2}\,g_{\mathbf q}\,,
		\qquad
		q^2\equiv \mathbf q^2\,.
	\end{equation}
	Therefore
	\begin{equation}\label{eq:J^LJ^L}
		\int_{\mathbf x}\, J_i^{L}J_i^{L}
		=
		\int_{\mathbf q}\frac{g_{\mathbf q}g_{-\mathbf q}}{q^2}\,,
		\qquad
		\int_{\mathbf q}\equiv \int \frac{d^3q}{(2\pi)^3}\,.
	\end{equation}
	Using $J^2=(J^L)^2+(J^T)^2$, Eq.~\eqref{eq:S_n} becomes
	\begin{equation}\label{eq:S[n,J]_expansion}
		S[n,J]
		=
		S[\langle n\rangle]
		-\frac{1}{2\chi_g}\int_{\mathbf q}\frac{g_{\mathbf q}g_{-\mathbf q}}{q^2}
		-\frac{1}{2\chi_g}\int_{\mathbf q} J_i^T(\mathbf q)J_i^T(-\mathbf q)\,.
	\end{equation}
	Since the transverse sector is completely decoupled from $n$ and $g$, integrating it out only changes the overall normalization. The reduced equilibrium measure in the variables $(n,g)$ is thus
	\begin{equation}\label{eq:P=e^S[n,g]}
		P_{\rm eq}[n,g]\propto e^{S[n,g]}\,,
	\end{equation}
	with
	\begin{equation}\label{eq:S_ng(q)}
		S[n,g]
		=
		S[\langle n\rangle]
		-\frac{1}{2\chi_g}\int_{\mathbf q}\frac{g_{\mathbf q}g_{-\mathbf q}}{q^2}\,.
	\end{equation}
	This equation is the equilibrium potential appropriate for the variables used in the main text.

	To make the density sector explicit, we expand the static thermodynamic functional around equilibrium in Eq.~\eqref{eq:alpha_expansion}. The equilibrium correlators of density fluctuation $\delta n$ are generated by the weight 
	\begin{equation}\label{eq:P=e^S[n]}
		P_n[n]\propto e^{S[n]}\,. 
	\end{equation} 
	Noting $W_N^{\rm eq}\equiv {\rm WT}[\langle \delta n_1\cdots \delta n_N\rangle_{c,{\rm eq}}]$, the first few equilibrium connected cumulants are given in Eq.~\eqref{eq:W234}.

	Because Eq.~\eqref{eq:S_ng(q)} also shows that the equilibrium measure factorizes:
	\begin{equation}\label{eq:P=PP}
		P_{\rm eq}[n,g]
		=
		P_n[n]\;P_g[g]\,,
	\end{equation}
	with
	\begin{equation}\label{eq:P_n-P_g}
		P_n[n]\propto e^{S[n]}\,,
		\qquad
		P_g[g]\propto
		\exp\!\left[
		-\frac{1}{2\chi_g}\int_{\mathbf q}\frac{g_{\mathbf q}g_{-\mathbf q}}{q^2}
		\right]\,.
	\end{equation}
	Hence the density sector is exactly the usual thermodynamic one, while the $g$ sector is purely Gaussian and statistically independent of $n$ in equilibrium.
	
	By definition we have $\langle \delta n\rangle_{\rm eq}=0$, $\langle \delta g\rangle_{\rm eq}=0$.
	Because the $g$ sector is Gaussian, all its equilibrium correlators follow immediately. The two-point function is
	\begin{equation}\label{eq:Y2_app}
		Y_{2}^{\rm eq}(q)\equiv {\rm WT}[G_{gg}^{\rm eq}]\equiv {\rm WT}[\langle\delta g_1\delta g_2\rangle_{\rm eq}]=
		\chi_g\,q^2\,,
	\end{equation}
	and all odd $g$ correlators vanish,
	\begin{equation}\label{eq:oddg}
		Y_3^{\rm eq}\equiv{\rm WT}[\langle \delta g_1\delta g_2\delta g_3\rangle_{\rm eq}]=0\,,
		\qquad
		Y_5^{\rm eq}\equiv{\rm WT}[\langle \delta g_1\delta g_2\delta g_3\delta g_4\delta g_5\rangle_{\rm eq}]=0\,,
		\qquad \ldots
	\end{equation}
	while all even ones are given by Wick's theorem, e.g.,
	\begin{equation}\label{eq:eveng}
		Y_4^{\rm eq}\equiv{\rm WT}[\langle\delta g_1\delta g_2\delta g_3\delta g_4\rangle_{\rm eq}]
		=
		Y_{12}Y_{34}+Y_{13}Y_{24}+Y_{14}Y_{23}\,.
	\end{equation}
	Since the $g$ sector is Gaussian, all connected $g$ cumulants above second order vanish:
	\begin{equation}\label{eq:nonGaussiang}
		Y_N^{\rm eq}\equiv\langle \delta g_1\dots\delta g_N\rangle_{c,{\rm eq}}=0\,.
	\end{equation}
	
	Because the equilibrium measure factorizes, any mixed correlator splits into an $n$ part and a $g$ part:
	\begin{equation}\label{eq:fact_correlators}
		\langle \mathcal O_n[n]\;\mathcal O_\pi[\pi]\rangle_{\rm eq}
		=
		\langle \mathcal O_n[n]\rangle_{\rm eq}
		\langle \mathcal O_\pi[\pi]\rangle_{\rm eq}\,.
	\end{equation}
	In particular,
	\begin{equation}\label{X_2_App}
		X_2^{\rm eq}\equiv{\rm WT}[\langle \delta g_1\delta n_2\rangle_{\rm eq}]=0
	\end{equation}
	and, more generally, all connected mixed cumulants vanish:
	\begin{equation}\label{eq:mixed_correlators}
		\langle \delta g\cdots \delta g\,\delta n\cdots \delta n\rangle_{c,{\rm eq}}=0
		\qquad
		\text{whenever both $n$ and $g$ are present.
		}
	\end{equation}
	The first few full mixed correlators are then 
	\begin{subequations}
		\begin{equation}
			\langle \delta n_1 \delta n_2\delta  g_3\rangle_{\rm eq}=0\,,
		\end{equation}
		\begin{equation}
			\langle \delta n_1 \delta g_2 \delta g_3\rangle_{\rm eq}
			=
			\langle\delta n_1\rangle_{\rm eq}\,Y_{23}=0\,,
		\end{equation}
		\begin{equation}
			\langle\delta  n_1\delta  n_2\delta  g_3 \delta g_4\rangle_{\rm eq}
			=
			\langle \delta n_1 \delta n_2\rangle_{\rm eq}\,
			\langle \delta g_3\delta g_4\rangle_{\rm eq}
			=
			W_2^{\rm eq}(12)\,Y_{34}\,,
		\end{equation}
		\begin{equation}
			\langle \delta n_1 \delta g_2 \delta g_3 \delta g_4\rangle_{\rm eq}=0\,,
		\end{equation}
	\end{subequations}
	and similarly at higher orders. Thus every equilibrium correlator containing an odd number of $g$'s vanishes, while those with an even number of $g$'s are obtained by multiplying the charge correlator by the Wick contractions of the $g$ sector.

	The equilibrium correlators derived above are precisely the stationary solutions of the evolution equations used in the main text, i.e. the solutions obtained by setting $\partial_t=0$ in the hierarchy for equal-time cumulants. At the two-point level, the evolution equations for the connected correlators
	\begin{equation}\label{eq:WXY_app}
		W_2(12)={\rm WT}[\langle \delta n_1 \delta n_2\rangle]\,,\qquad X_2(12)={\rm WT}[\langle \delta g_1\delta n_2\rangle]\,,\qquad Y_2(12)={\rm WT}[\langle \delta g_1\delta g_2\rangle] 
	\end{equation} 
	admit the stationary solution 
	\begin{equation}\label{eq:WXY_eq_app}
		W_2^{\rm eq}=\frac{1}{\alpha'}\,, \qquad X_2^{\rm eq}=0\,, \qquad  Y_2^{\rm eq}(q) =\frac{T\lambda}{\tau}\,q^2\,, 
	\end{equation}
	in complete agreement with Eqs.~\eqref{eq:W234}, \eqref{X_2_App} and \eqref{eq:Y2_app}. Likewise, at higher orders one finds 
	\begin{equation}\label{eq:mixed_correlators_eq}
		\langle\delta g\cdots \delta g\,\delta n\cdots \delta n\rangle_{c,{\rm eq}}=0 \quad \text{for all mixed connected sectors,}
	\end{equation}
	while the purely density connected correlators are exactly the static thermodynamic cumulants generated by $S[n]$, namely $W_3^{\rm eq}$, $W_4^{c,{\rm eq}}$, and so on, as given above. Thus, in the constant-coefficient case relevant for the paper, the Gibbs measure in the $(n,g)$ variables reproduces exactly the $\partial_t=0$ limit of the dynamical hierarchy.

	\section{Numerical implementation}
	\label{app:numerical}
	
	We outline here the numerical strategy for integrating the linear $W_2$ system. The concrete inputs
	for $\gamma(t)$, $\chi(t)$, and $\tau(t)$ (and the associated equation of state) will be specified in the next section,
	where we implement this strategy for phenomenological simulations.
	
	\emph{Why do we not discretize the stochastic equations in real space and simulate noise realizations?}
	The observables of interest---cumulants within a finite acceptance---are determined by connected
	equal-time correlators. It is therefore far more efficient to evolve these correlators (Wigner functions)
	directly, rather than generating noise on a lattice and averaging over many events. This bypasses the need
	for large event statistics and eliminates Monte-Carlo noise.
	
	Moreover, a real-space discretization of the stochastic MC system would require resolving
	both the slow diffusive scale and the microscopic relaxation scale $\tau$, and it would introduce ultraviolet
	sensitivity associated with white noise at the lattice scale. By working instead with the closed evolution
	equations for $W_2$ (and higher $W_N$), these issues are absorbed into controlled input functions and
	renormalized coefficients (e.g.\ $\gamma(t)$, $\chi(t)$, $\tau(t)$), so the numerical task reduces to integrating a
	deterministic system of ordinary differential equations for each Fourier mode.
	
	To start, let us take the real vector $U$:
	\begin{equation}\label{eq:U^(2)}
		U^{(2)}=\big(W_2,\,X_r,\,X_i,\,Y_2\big)^T\equiv\,\Big(W_2(\mathbf q),\,\mathrm{Re} X_2(\mathbf q),\, \mathrm{Im} X_2(\mathbf q),\,Y_2(\mathbf q)\Big)^T\,.
	\end{equation}
	Then the system of equations becomes
	\begin{eqnarray}\label{eq:evo_WXY2}\nonumber
		\partial_t W_2 &=& 2X_r\,,\\\nonumber
		\partial_t X_r &=& -\frac{\gamma q^2}{\tau} W_2 - \frac{1}{\tau} X_r + Y_2\,,
		\qquad\\\nonumber
		\partial_t X_i &=& -\frac{1}{\tau} X_i\,,\\
		\partial_t Y_2 &=& -\frac{2\gamma q^2}{\tau} X_r - \frac{2}{\tau} Y_2 
		+ \frac{2\lambda}{\tau^2} q^2\,.
	\end{eqnarray}
	which can be also simply written as
	\begin{equation}\label{eq:dtU^(2)}
		\partial_t U^{(2)} = A^{(2)}\,U^{(2)} + S^{(2)},
	\end{equation}
	with
	\begin{equation}\label{eq:AS^(2)}
		A^{(2)} =
		\begin{pmatrix}
			0 & 2 & 0 & 0 \\[8pt]
			-\dfrac{\gamma q^2}{\tau} & -\dfrac{1}{\tau} & 0 & 1 \\[10pt]
			0 & 0 & -\dfrac{1}{\tau} & 0 \\[10pt]
			0 & -\dfrac{2\gamma q^2}{\tau} & 0 & -\dfrac{2}{\tau}
		\end{pmatrix},
		\qquad
		S^{(2)} =
		\begin{pmatrix}
			0 \\[4pt]
			0 \\[4pt]
			0 \\[4pt]
			\dfrac{2\lambda}{\tau^2} q^2
		\end{pmatrix}.
	\end{equation}	
	We may integrate in time using implicit Euler with a uniform step $\Delta t$:
	\begin{equation}\label{eq:disc_dtU^(2)}
		\bigl(I - \Delta t\, A^{(2)}\bigr)\, U^{(2)}_{\,n+1}
		= U^{(2)}_{n} + \Delta t\, S^{(2)}(t_{n+1})\,.
	\end{equation}		
	Since $A$ is constant, the matrix $\bigl(I - \Delta t\, A\bigr)$ is also constant for a given triangle and chosen $\Delta t$. 
	A practical choice is to take $\Delta t$ small compared to the fastest linear relaxation time.
	In MC theory the linear spectrum contains two branches, obtained from the dispersion relation
	\begin{equation}\label{eq:MC_dispersion}
		\tau\,\omega^2- i \omega + \gamma  q^2 =0\,,
	\end{equation}
	so that
	\begin{equation}\label{eq:omega_pm}
		\omega_{\pm}(q)=\frac{-i\pm i\sqrt{1-4\tau \gamma q^2}}{2\tau}\,.
	\end{equation}
	The slow (hydrodynamic) branch is the one that reduces to diffusion at small $q$, i.e., $\omega_{+}$, and we define the corresponding relaxation time by
	\begin{equation}\label{eq:diff_time}
		\tau_\gamma(q)\equiv \frac{1}{|\omega_{\rm slow}(q)|}\,.
	\end{equation}
	Accordingly, we choose
	\begin{equation}\label{eq:Deltat_domain}
		\Delta t \ll \min\!\left(\tau,\;\tau_\gamma(q_{\max})\right)\,.
	\end{equation}	
	In practice, when constructing the set of momentum triangles to be evolved and later integrated over
	acceptance, we restrict to \(|q|\le q_{\rm cut}\), where \(q_{\rm cut}\) is the ultraviolet cutoff of the
	effective description (set by coarse graining and/or by the acceptance kernel).	
	\paragraph{Initial conditions.}
	To start the evolution at $t=t_0$ we must specify $U(t_0)$, i.e.
	$W(t_0)$, $X_r(t_0)$, $X_i(t_0)$ and $Y(t_0)$.
	Physically, $W$ is the equal-time charge correlator, while $X$ and $Y$ encode
	correlators involving the auxiliary variable $g\equiv\partial_t n$.
	In particular, in the diffusive limit $\tau\to 0$ one has $g=\partial_t n$ becoming a
	distribution (time derivative of a noisy process), so equal-time objects such as
	$\langle gg\rangle$ (and hence $Y$) become parametrically large. This is why
	the $\tau\to 0$ limit is singular at the level of $U$ even though $W$ has a smooth
	diffusive limit.
	
	In practice we use the following consistent initialization strategy.
	First, we always set
	\begin{equation}\label{eq:ICX}
		X_i(t_0)=0\,,
	\end{equation}
	since $X_i$ is completely decoupled and simply decays as $e^{-(t-t_0)/\tau}$.
	For $X_r$ we take
	\begin{equation}\label{eq:ICXr}
		X_r(t_0)=0\,,
	\end{equation}
	which corresponds to the absence of an initial ``velocity--density'' correlation.
	
	Next, $W_2$ is initialized either in local equilibrium:
	\begin{equation}\label{eq:ICW}
		W_2(t_0)=W_2^{\rm eq}= \frac{1}{\alpha'(t_0)}\,.
	\end{equation}
	Finally, $Y_2$ must be chosen in a way compatible with the fast MC sector.
	A convenient and numerically stable choice is to initialize $Y_2$ at its
	quasi-stationary (fast-mode) value at fixed $W_2$ and $X_r$, obtained by imposing
	$\partial_t Y_2(t_0)=0$ in Eq.~\eqref{eq:evo_Y2}. 
	With Eq.~\eqref{eq:ICXr}, this reduces to
	\begin{equation}\label{eq:ICYsimple}
		Y_2(t_0)=\frac{\lambda(t_0)}{\tau(t_0)}\,q^2\,.
	\end{equation}
	This choice ensures that the large source term $\sim 2\lambda q^2/\tau^2$ in $\partial_t Y_2$
	is balanced from the start, avoiding an artificial $\mathcal O(\tau^{-1})$ transient
	in the auxiliary correlator. Importantly, the charge correlator $W_2$ remains finite
	and smoothly approaches the diffusion result as $\tau\to 0$.

	For the special initialization above, one indeed has $\partial_t U(t_0)=0$.
	When we intentionally initialize $W_2(t_0)$ away from $W_2^{\rm eq}$, $\partial_t U(t_0)\neq 0$ by construction and the system
	relaxes toward the instantaneous equilibrium correlator.

	\subsection{Three-point functions}
	\label{subapp:W3}
	
	For a momentum triangle $(q_1, q_2, q_3)$ with $	q_1 + q_2 + q_3 = 0,$
	we	define
	\begin{equation}\label{eq:ng_3pt}
		n_a \equiv n_{q_a}\,, \qquad g_a \equiv g_{q_a} = \partial_t n_{q_a}\,, \qquad a \in \{1,2,3\}\,.
	\end{equation}
	Let us also define the $(3)$-sector equal-time correlators where $W,X,Y,Z$ represents three-point correlators with 0, 1, 2, 3 $g$-field insertions, respectively:
	\begin{equation}\label{eq:def_WXYZ3_app}
		\begin{split}
			W_3 &\equiv {\rm WT}[\langle n_1 n_2 n_3\rangle]\,,\\
			X_a^{(3)} &\equiv{\rm WT}[\langle g_a\, n_b n_c\rangle]\,, \qquad\\
			Y_{ab}^{(3)} &\equiv {\rm WT}[\langle g_a g_b\, n_c\rangle] \quad (a<b)\,,\\
			Z^{(3)} &\equiv {\rm WT}[\langle g_1 g_2 g_3\rangle]\,,
		\end{split}
	\end{equation}
	where $\{a,b,c\}=\{1,2,3\}$.
	Note that $X^{(3)}_a$ is the ``one–$g$'' insertion on leg $a$; $Y^{(3)}_{ab}$ is the ``two–$g$'' insertion on legs $a,b$.
	
	\paragraph{Evolution equations and the nonlinear terms.}	We find that the set of equations is given by 
	\begin{align}\label{eq:W_3_W}
		\partial_t W_3&= X_1^{(3)} + X_2^{(3)} + X_3^{(3)}\,,\\ \label{X_3_W}
		\tau\, \partial_t X_a^{(3)} &= -X_a^{(3)} - \gamma q_a^2 W_3^{(3)}
		+ \tau \sum_{b\neq a} Y_{ab}^{(3)} + S_a^{(3)}\,,\\\label{Y_3_W}
		\tau\, \partial_t Y_{ab}^{(3)}&= -2 Y_{ab}^{(3)}
		- \gamma\left(q_a^2 X_b^{(3)} + q_b^2 X_a^{(3)}\right)
		+ \tau\, Z^{(3)}+\, S_{ab}^{(3)}\,,\\\label{Z_3_W}
		\tau\, \partial_t Z^{(3)}&= -3 Z^{(3)}
		- \gamma\left(q_1^2 Y_{23}^{(3)} + q_2^2 Y_{13}^{(3)} + q_3^2 Y_{12}^{(3)}\right)+\,S^{(3)}\,.
	\end{align}
	where the source terms which drive the ODE are given by 
	\begin{align}\label{eq:S^(3)}
		S_a^{(3)} &= -\gamma'\, q_a^2\, W_2(q_b)\, W_2(q_c)\,, \qquad \{a,b,c\}=\{1,2,3\}\,,\\
		S_{ab}^{(3)}&=-\gamma'\Big[
		q_a^2X_2(q_b)W_2(q_c)
		+q_b^2X_2(q_a)W_2(q_c)
		\Big]\,,\\
		S^{(3)}&=	-\gamma'\!\left[
		q_1^2 X_2(q_2)X_2(q_3)
		+
		q_2^2 X_2(q_1)X_2(q_3)
		+
		q_3^2 X_2(q_1)X_2(q_2)
		\right]\,.
	\end{align}
	
	\paragraph{Matrix form.}	Equations \eqref{eq:W_3_W}--\eqref{Z_3_W} can be written as the following 
	\begin{equation}\label{eq:dtU^(3)}
		\partial_t U^{(3)} = A^{(3)}\, U^{(3)} + b^{(3)}\,,
	\end{equation}
	with
	\begin{equation}\label{eq:U^(3)}
		U^{(3)} \equiv 
		\left(
		W_3,\;
		X_1^{(3)},\;
		X_2^{(3)},\;
		X_3^{(3)},\;
		Y_{12}^{(3)},\;
		Y_{13}^{(3)},\;
		Y_{23}^{(3)},\;
		Z^{(3)}
		\right)^{T}\,.
	\end{equation}
	and 
	\begin{equation}\label{eq:b^(3)}
		b^{(3)} = 
		\left(
		0,\;
		\frac{S_1^{(3)}}{\tau},\;
		\frac{S_2^{(3)}}{\tau},\;
		\frac{S_3^{(3)}}{\tau},\;
		\frac{S_{12}^{(3)}}{\tau},\;\frac{S_{13}^{(3)}}{\tau},\;\frac{S_{23}^{(3)}}{\tau},\;\frac{S^{(3)}}{\tau}
		\right)^{T}\,.
	\end{equation}
	In addition, $A^{(3)}$ is the constant $8\times 8$ matrix read off directly from the coefficients in the ODEs above (depending only on $\tau,\gamma,q_1^2,q_2^2,q_3^2$).
	More precisely, we may let $a=q_1^2$, $b=q_2^2$, $c=q_3^2$. Then the constant matrix reads
	\begin{equation}\label{eq:A^(3)}
		A^{(3)} =
		\begin{pmatrix}
			0 & 1 & 1 & 1 & 0 & 0 & 0 & 0 \\[4pt]
			-\gamma a/\tau & -1/\tau & 0 & 0 & 1 & 1 & 0 & 0 \\[4pt]
			-\gamma b/\tau & 0 & -1/\tau & 0 & 1 & 0 & 1 & 0 \\[4pt]
			-\gamma c/\tau & 0 & 0 & -1/\tau & 0 & 1 & 1 & 0 \\[4pt]
			0 & -\gamma b/\tau & -\gamma a/\tau & 0 & -2/\tau & 0 & 0 & 1 \\[4pt]
			0 & -\gamma c/\tau & 0 & -\gamma a/\tau & 0 & -2/\tau & 0 & 1 \\[4pt]
			0 & 0 & -\gamma c/\tau & -\gamma b/\tau & 0 & 0 & -2/\tau & 1 \\[4pt]
			0 & 0 & 0 & 0 & -\gamma c/\tau & -\gamma b/\tau & -\gamma a/\tau & -3/\tau
		\end{pmatrix}\,.
	\end{equation}		
	We integrate Eq.~\eqref{eq:dtU^(3)} in time using implicit Euler with a uniform step $\Delta t$:
	\begin{equation}\label{eq:disc_dtU^(3)}
		\bigl(I - \Delta t\, A^{(3)}\bigr)\, U^{(3)}_{\,n+1}
		= U^{(3)}_{n} + \Delta t\, b^{(3)}(t_{n+1})\,.
	\end{equation}		
	Since $A^{(3)}$ is constant, the matrix $\bigl(I - \Delta t\, A^{(3)}\bigr)$ is also constant for a given triangle and chosen $\Delta t$. 
	We choose
	\begin{equation}\label{eq:Deltat_domain3}
		\Delta t \ll \min\!\left(\tau,\;\tau_\gamma(q_{\max})\right)\,,
		\qquad
		q_{\max}\equiv \max\!\left(|q_1|,|q_2|,|q_3|\right)\,.
	\end{equation}	
	See \eqref{eq:diff_time} for the defintion of $\tau_\gamma(q)$.	In practice, when constructing the set of momentum triangles to be evolved and later integrated over
	acceptance, we restrict to \(|q_a|\le q_{\rm cut}\), where \(q_{\rm cut}\) is the ultraviolet cutoff of the
	effective description (set by coarse graining and/or by the acceptance kernel).
	
	\paragraph{Initial conditions.}
	For each momentum triangle $(q_1,q_2,q_3)$ we initialize the $W_3$ sector by imposing instantaneous
	stationarity at $t=t_0$, \emph{i.e.}\ we set
	\begin{equation}\label{eq:IC_Udot0_W3}
		\partial_t U^{(3)}(t_0)=0
		\qquad\Longleftrightarrow\qquad
		A^{(3)}(t_0)\,U^{(3)}(t_0)+b^{(3)}(t_0)=0,
	\end{equation}
	with $b^{(3)}(t)$ determined by the source terms
	$S_a^{(3)}(t)=-\gamma' q_a^2\,W_2(q_b,t)\,W_2(q_c,t)$.
	Operationally, we solve the $8\times 8$ linear system \eqref{eq:IC_Udot0_W3} once for each triangle at $t_0$.
	This choice determines all components of $U^{(3)}(t_0)$ self-consistently: once the source $b^{(3)}(t_0)$
	is fixed by $W_2(t_0)$, the condition $\partial_t U^{(3)}(t_0)=0$ fixes the charge correlator
	$W_3^{(3)}(q_1,q_2,q_3;t_0)$ and simultaneously sets the auxiliary $g$-inserted correlators
	$X_a^{(3)}(t_0)$, $Y_{ab}^{(3)}(t_0)$ and $Z^{(3)}(t_0)$ to their instantaneous steady-state values.
	In this way we avoid an artificial short-time transient on the fast relaxation scale $\sim \tau$ that would
	otherwise appear if these auxiliary correlators were initialized by hand.
	
	Let us emphasize that imposing $\partial_t U^{(3)}(t_0)=0$ only fixes the \emph{instantaneous} stationary state at $t_0$; for $t>t_0$ the
	system is driven out of this state because the coefficients and the source $b^{(3)}(t)$ evolve in time, via $T(t)$, $\mu(t)$ and the time-dependent two-point function $W_2$.
	Let us recall that in our framework the time dependence enters through the evolving background $T(t),\mu(t)$, which determines
	the coefficients in $A^{(n)}(t)$ and the instantaneous equilibrium correlators.  The two-point function
	$W_2$ is driven toward $W_2^{\rm eq}(t)$ by the noise–dissipation balance (FDT), while higher-point
	correlators are generated by deterministic source terms built from lower correlators (e.g.\ $S^{(3)}\sim
	\gamma'(t)\,W_2 W_2$), whose magnitude is fixed by the equation of state. We will discuss the equation of state in the next section.

	\subsection{Four-point functions}
	\label{subapp:W4}
	
	We first fix a momentum quadruplet (a “quadrilateral”) $	q_1+q_2+q_3+q_4=0\,.$
	We define the equal-time connected four-point correlators where $W,X,Y,Z,R$ represents four-point correlators with 0, 1, 2, 3, 4 $g$-field insertions, respectively:
	\begin{equation}\label{eq:def_WXYZR4_app}
		\begin{split}
			W^c_4 &\equiv {\rm WT}[\langle n_1 n_2 n_3 n_4 \rangle_c]\,,\\
			X_a^{(4)} &\equiv {\rm WT}[\Big\langle g_a \prod_{b\neq a} n_b \Big\rangle_c]\,,\\
			Y_{ab}^{c,(4)} &\equiv {\rm WT}[\Big\langle g_a g_b \prod_{c\neq a,b} n_c \Big\rangle_c]\,,\qquad a<b\,,\\
			Z_{abc}^{c,(4)} &\equiv {\rm WT}[\langle g_a g_b g_c n_d \rangle_c]\,, 
			\qquad \text{where } \{a,b,c,d\}=\{1,2,3,4\}\,,\\
			R^{c,(4)} &\equiv {\rm WT}[\langle g_1 g_2 g_3 g_4 \rangle_c]\,.
		\end{split}
	\end{equation}
	A convenient 16-component state vector is then
	\begin{equation}\label{eq:U^(4)}
		U^{c,(4)}=
		\big(
		W^c_4,\;
		\underbrace{X_a^{c,(4)}}_4,\;
		\underbrace{Y_{ab}^{c,(4)}}_6,\;
		\underbrace{Z_{abc}^{c,(4)}}_4,\;
		R^{c.(4)}
		\big)^{T}\,.
	\end{equation}
	\paragraph{Evolution equations.}
	For fixed $\tau$, $\gamma$ and fixed $\{q_a\}$, the $W_4$ sector obeys a closed linear system:
	\begin{align}
		\frac{d}{dt} W_4^c
		&= X_1^{c,(4)}+X_2^{c,(4)}+X_3^{(4)}+X_4^{c,(4)}\,, \label{eq:W4_W}\\
		\tau\,\frac{d}{dt} X_a^{(4)}
		&= -X_a^{c,(4)}-\gamma\,q_a^2\,W_4^{c,(4)}
		+\tau\sum_{b\neq a} Y_{ab}^{c,(4)} + S_a^{c,(X,4)}(t)\,, \label{eq:W4_X}\\
		\tau\,\frac{d}{dt} Y_{ab}^{(4)}
		&= -2\,Y_{ab}^{c,(4)}
		-\gamma\!\left(q_a^2 X_b^{c,(4)}+q_b^2 X_a^{c,(4)}\right)
		+\tau\!\left(Z_{abc}^{c,(4)}+Z_{abd}^{c,(4)}\right)+ S_{ab}^{c,(Y,4)}\,, \label{eq:W4_Y}\\
		\tau\,\frac{d}{dt} Z_{abc}^{c,(4)}
		&= -3\,Z_{abc}^{c,(4)}
		-\gamma\!\left(
		q_a^2 Y_{bc}^{c,(4)}+q_b^2 Y_{ac}^{c,(4)}+q_c^2 Y_{ab}^{c,(4)}
		\right)
		+\tau\,R^{c,(4)}+ S_{abc}^{c,(Z,4)}\,, \label{eq:W4_Z}\\
		\tau\,\frac{d}{dt} R^{c,(4)}
		&= -4\,R^{c,(4)}
		-\gamma\!\left(
		q_1^2 Z_{234}^{c,(4)}+
		q_2^2 Z_{134}^{c,(4)}+
		q_3^2 Z_{124}^{c,(4)}+
		q_4^2 Z_{123}^{c,(4)}
		\right)+ S^{c,(R,4)}\,. \label{eq:W4_R}
	\end{align}
	\paragraph{Nonliear terms.}
	The nonliear term splits into the contributions from the quadratic and cubic vertices,
	The $\gamma'$ term arises from the quadratic vertex  while the $\gamma''$ term arises from the cubic vertex. Each $W_3$ argument forms a triangle because $(q_a+q_b)+q_c+q_d=0$. We will take $q_{ab}\equiv q_a+q_b$ in the following. 
	We find, for $\{a,b,c,d\}=\{1,2,3,4\}$,
	\begin{equation}
		\begin{split}
			S_a^{c,(X,4)}	=&
			-\gamma' q_a^2\Big[
			W_2(q_b)\,W_3(q_{ab},q_c,q_d)
			+
			W_2(q_c)\,W_3(q_{ac},q_b,q_d)
			+
			W_2(q_d)\,W_3(q_{ad},q_b,q_c)
			\Big]\\	&
			-3\gamma'' q_a^2\,W_2(q_b)W_2(q_c)W_2(q_d)\,,
		\end{split}
	\end{equation}
	with  $\{c,d\}=\{1,2,3,4\}\setminus\{a,b\}$\footnote{This takes the set $\{1,2,3,4\}$, and removes the elements $a$ and $b$. So if, for example, $a=1$ and $b=3$, then $\{1,2,3,4\}\setminus\{a,b\}=\{2,4\}$.},
	\begin{align}
		S_{ab}^{(Y,4)}
		={}&
		-\gamma' q_a^2\Big[
		W_2(q_c)\,X_b^{(3)}(q_{ac},q_d)
		+
		W_2(q_d)\,X_b^{(3)}(q_{ad},q_c)
		\color{black}	+
		X_2(q_b)\,W_3(q_{ab},q_c,q_d)
		\Big]
		\nonumber\\
		&-\gamma' q_b^2\Big[
		W_2(q_c)\,X_a^{(3)}(q_{bc},q_d)
		+
		W_2(q_d)\,X_a^{(3)}(q_{bd},q_c)
		\color{black}+
		X_2(q_a)\,W_3(q_{ab},q_c,q_d)
		\Big]
		\nonumber\\
		&-\gamma''\Big[
		q_a^2\,W_2(q_c)W_2(q_d)X_2(q_b)
		+
		q_b^2\,W_2(q_c)W_2(q_d)X_2(q_a)
		\Big]\,.
	\end{align}
	Note also that with $q_a+q_b+q_c=0 $:
	\begin{align}
		X_{a}^{(3)}&\equiv{\rm WT}[\langle g_{q_a} n_{q_b}n_{q_c}\rangle]\equiv\, X_{a}^{(3)}(q_b,q_c)\,,\nonumber\\
		Y_{ab}^{(3)}&\equiv{\rm WT}[\langle g_{q_a} g_{q_b}n_{q_c}\rangle]\equiv\,Y_{ab}^{(3)}(q_c)\,.
	\end{align}
	With this and $\{b,c,d\}=\{1,2,3,4\}\setminus\{a\}$, we find, with $q_a+q_b+q_c+q_d=0$,
	\begin{equation}
		\begin{aligned}
			S_{abc}^{c,(Z,4)}
			={}&-\gamma'\Big[
			q_a^2 W_2(q_d)\,	Y^{(3)}_{bc}(q_{ad})+
			q_b^2 W_2(q_d)\,Y^{(3)}_{ac}(q_{bd})
			+	q_c^2 W_2(q_d)\,	Y^{(3)}_{ab}(q_{cd}) \color{black}\Big]
			\\[1mm]
			&-\gamma'\Big[
			\Big(q_a^2 X_2(q_b)+q_b^2 X_2(q_a)\Big)\,X_c^{(3)}(q_d,q_{ab})
			+ \Big( q_b^2 X_2(q_c)+q_c^2 X_2(q_b)\Big)\,X_a^{(3)}(q_d,q_{bc})
			\\
			&\hspace{1.35cm}
			+\Big(q_c^2  X_2(q_a)+q_a^2 X_2(q_c)\Big)\,X_b^{(3)}(q_d,q_{ac})
			\Big]
			\\[1mm]
			&-\gamma''\Big[
			q_a^2 X_2(q_b)X_2(q_c)W_2(q_d)
			+q_b^2 X_2(q_c)X_2(q_a)W_2(q_d)
			+q_c^2 X_2(q_a)X_2(q_b)W_2(q_d)
			\Big]\,,
		\end{aligned}
	\end{equation}
	with $d$ the remaining leg. Finally we have
	\begin{align}
		S^{(R,4)}
		={}&
		-\gamma' q_a^2\Big[
		X_2(q_b)\,Y^{(3)}_{cd}(q_{ab})
		+
		X_2(q_c)\,Y^{(3)}_{bd}(q_{ac})
		+
		X_2(q_d)\,Y^{(3)}_{bc}(q_{ad})
		\Big]
		\nonumber\\
		&-\gamma' q_b^2\Big[
		X_2(q_a)\,Y^{(3)}_{cd}(q_{ab})
		+
		X_2(q_c)\,Y^{(3)}_{ad}(q_{bc})
		+
		X_2(q_d)\,Y^{(3)}_{ac}(q_{bd})
		\Big]
		\nonumber\\
		&-\gamma' q_c^2\Big[
		X_2(q_a)\,Y^{(3)}_{bd}(q_{ac})
		+
		X_2(q_b)\,Y^{(3)}_{ad}(q_{bc})
		+
		X_2(q_d)\,Y^{(3)}_{ab}(q_{cd})
		\Big]
		\nonumber\\
		&-\gamma' q_d^2\Big[
		X_2(q_a)\,Y^{(3)}_{bc}(q_{ad})
		+
		X_2(q_b)\,Y^{(3)}_{ac}(q_{bd})
		+
		X_2(q_c)\,Y^{(3)}_{ab}(q_{cd})
		\Big]
		\nonumber\\\nonumber
		&-\gamma''\Big[
		q_a^2 X_2(q_b)X_2(q_c)X_2(q_d)
		+
		q_b^2 X_2(q_a)X_2(q_c)X_2(q_d)\\
		&
		\qquad\qquad\quad+
		q_c^2 X_2(q_a)X_2(q_b)X_2(q_d)
		+
		q_d^2 X_2(q_a)X_2(q_b)X_2(q_c)
		\Big]\,.
	\end{align}
	As a result, the $W_4$ sector is a linear ODE system driven by known lower-sector inputs
	$W_2(t)$ and $W_3(t)$:
	\begin{equation}
		\frac{d}{dt}U^{(4)}=A^{(4)}\,U^{(4)}+b^{(4)}\,,
	\end{equation}
	where $A^{(4)}$ is a constant $16\times16$ matrix (for fixed $\tau$, $\gamma$, and fixed $\{q_a\}$), and
	\begin{align}
		b^{(4)}(t)=\Big(
		&0,\,
		S_1^{(X,4)}(t)/\tau,\,
		S_2^{(X,4)}(t)/\tau,\,
		S_3^{(X,4)}(t)/\tau,\,
		S_4^{(X,4)}(t)/\tau,\,
		\nonumber\\
		& S_{12}^{(Y,4)}(t)/\tau,\,
		S_{13}^{(Y,4)}(t)/\tau,\,
		S_{14}^{(Y,4)}(t)/\tau,\,
		S_{23}^{(Y,4)}(t)/\tau,\,
		S_{24}^{(Y,4)}(t)/\tau,\,
		S_{34}^{(Y,4)}(t)/\tau,\,
		\nonumber\\
		& S_{123}^{(Z,4)}(t)/\tau,\,
		S_{124}^{(Z,4)}(t)/\tau,\,
		S_{134}^{(Z,4)}(t)/\tau,\,
		S_{234}^{(Z,4)}(t)/\tau,\,
		S^{(R,4)}(t)/\tau
		\Big)^T \,.
	\end{align}	
	The initialization of $U^{(4)}$ follows the same procedure as in the $W_3$ sector and will not be repeated here.

	\bibliography{References}
	
\end{document}